\documentclass[preprint,12pt]{elsarticle}

\usepackage{amssymb}

\usepackage{listings}
\usepackage{xcolor,colortbl}
\usepackage{multirow}
\usepackage{lscape}
\usepackage{rotating}
\usepackage{amsmath,amssymb,amsfonts}
\usepackage{graphicx}
\usepackage{textcomp}
\usepackage{xcolor}
\usepackage{enumitem}
\usepackage{listings}
\usepackage{adjustbox}

\usepackage{color}
\usepackage{multirow}
\usepackage{pdflscape}
\definecolor{dkgreen}{rgb}{0,0.6,0}
\definecolor{gray}{rgb}{0.5,0.5,0.5}
\definecolor{mauve}{rgb}{0.58,0,0.82}

\definecolor{Gray}{gray}{0.85}
\definecolor{LightCyan}{rgb}{0.88,1,1}

\newcolumntype{a}{>{\columncolor{Gray}}c}
\newcolumntype{b}{>{\columncolor{white}}c}
\usepackage{xcolor}

\definecolor{codegreen}{rgb}{0,0.6,0}
\definecolor{codegray}{rgb}{0.5,0.5,0.5}
\definecolor{codepurple}{rgb}{0.58,0,0.82}
\definecolor{backcolour}{rgb}{0.95,0.95,0.92}

\lstdefinestyle{mystyle}{
    backgroundcolor=\color{backcolour},   
    commentstyle=\color{codegreen},
    keywordstyle=\color{magenta},
    numberstyle=\tiny\color{codegray},
    stringstyle=\color{codepurple},
    basicstyle=\ttfamily\footnotesize,
    breakatwhitespace=false,         
    breaklines=true,                 
    captionpos=b,                    
    keepspaces=true,                 
    numbers=left,                    
    numbersep=5pt,                  
    showspaces=false,                
    showstringspaces=false,
    showtabs=false,                  
    tabsize=2
}

\begin{document}

\begin{frontmatter}



\title{Effective Fault Localization using Probabilistic and Grouping Approach}

\author[inst1]{Saksham Sahai Srivastava}

\affiliation[inst1]{organization={Dept. of Chemical Engineering},
            addressline={IIT Kharagpur}, 
            city={Kharagpur},
            postcode={721302}, 
            country={India}}

\author[inst2]{Arpita Dutta}

\affiliation[inst2]{organization={School of Computing},
            addressline={National University of Singapore}, 
            postcode={117417}, 
            country={Singapore}}

\author[inst3]{Rajib Mall}

\affiliation[inst3]{organization={Dept. of Computer Science \& Engineering},
            addressline={IIT Kharagpur}, 
            city={Kharagpur},
            postcode={721302}, 
            country={India}}

\begin{abstract}
\noindent\textbf{Context:} Fault localization (FL) is the key activity while debugging a program. Any improvement to this activity leads to significant improvement in total software development cost. There is an internal linkage between the program spectrum and test execution result. Conditional probability in statistics captures the probability of occurring one event  in relationship to one or more other events.  \\ 
\noindent\textbf{Objectives:}  The aim of this paper is to use the conception of conditional probability to design an effective fault localization technique. \\
\noindent\textbf{Methods:}  In the paper, we present a fault localization technique that derives the association between statement coverage information and test case execution result using condition probability statistics. This association with the failed test case result shows the fault containing probability of that specific statement. Subsequently, we use a grouping method to refine the obtained statement ranking sequence for better fault localization. \\
\noindent\textbf{Results:} We evaluated the effectiveness of proposed method over \textcolor{black}{ eleven open-source data sets}. Our obtained results show that on average, the proposed CGFL method is 24.56\% more effective than other contemporary fault localization methods such as D$^*$, Tarantula, Ochiai, Crosstab, BPNN, RBFNN, DNN, and CNN. \\
\noindent\textbf{Conclusion:} We devised an effective fault localization technique by combining the conditional probabilistic method with failed test case execution-based approach. Our experimental evaluation shows our proposed method outperforms the existing fault localization techniques.
\end{abstract}

%


\end{frontmatter}


\section{Introduction}
\label{sec1}
\par Software development comprises two immensely important as well as labor-intensive components of software testing and debugging. With the advancement in software development, softwares are growing in scale as well as in complexity, due to which execution failure can be considered inevitable. During software maintenance\cite{wong2016survey}, software debugging turns out to be the most arduous task. Software debugging comprises two major activities: localization of fault and the rectification of the fault. Fault localization (FL) is the process of uncovering faults present in different regions of the program which have been responsible for execution failures. There have been multiple attempts to design automated techniques which can achieve the objective of fault localization. Any improvement to these pre-existing automated techniques would significantly reduce total software maintenance cost and also debugging time\cite{mall2018fundamentals}.
\par In the past two-to-three decades, several automated FL techniques  have been reported   for effective localization of faults  with reduced human intervention. These proposed approaches aimed at localizing the faults by examining a small fraction of the code because the lesser the code would be examined, the better would be the effectiveness of the technique.  The fault localization techniques can be broadly classified into these categories:  slicing-based \cite{weiser1984program,korel1988dynamic}, spectrum-based (SBFL) techniques \cite{wong2008crosstab,jones2002visualization,wong2010family,wong2013dstar}, machine learning-based \cite{ascari2009exploring,dutta2019effective,wong2009bp,wong2011effective}, and  mutation-based techniques\cite{papadakis2015metallaxis,moon2014ask,dutta2021msfl,combifl}. There are three types of slicing techniques: static\cite{weiser1984program},  dynamic\cite{korel1988dynamic}, and execution slice-based methods\cite{agrawal1995fault}. Slicing-based techniques worked predominantly on the principle of deleting irrelevant segments of the program, thereby making the remainder of the program behave in a similar fashion as previously(without deletion) with respect to certain specifications. Spectrum-Based Fault Localization (SBFL) techniques\cite{jones2005empirical,wong2013dstar} by and large require program spectra information to give quantitative knowledge for localizing faults in a program. The program spectra include statement coverage information (executed/not executed) and test case execution results (pass/fail) which are supplied as input to the fault localizer. The fault localizer is expected to generate a ranked list of statements based on their suspicious scores of containing a bug. Machine learning which helps formulate mathematical models for complex problems also finds its application in solving the problem of fault localization. Therefore,  there are machine learning based techniques such as SVM\cite{ascari2009exploring}, ensemble classifier\cite{dutta2019effective}, decision trees\cite{briand2007using} and neural network models\cite{wong2009bp,wong2011effective,xiao2021albfl,li2021fault,lou2021boosting} which manifested promising results for FL. The mutation based techniques such Metallaxis\cite{papadakis2015metallaxis}, MUSE\cite{moon2014ask}, MSFL\cite{dutta2021msfl} and Combi-FL\cite{combifl} have also presented substantial improvement in localizing faults effectively.   
\par But these antecedent approaches hold certain drawbacks. Slicing-based techniques\cite{weiser1984program,korel1988dynamic} possess the disadvantage of not assigning ranks to the program statements while the SBFL techniques\cite{jones2005empirical,naish2011model} generate a ranked list with many statements having tie in ranks. The machine learning-based techniques\cite{ascari2009exploring,dutta2019effective,wong2009bp,wong2011effective,xiao2021albfl,li2021fault,lou2021boosting} although being effective, require a plenty amount of time to complete the computation and therefore have low efficiency. The mutation-based techniques\cite{papadakis2015metallaxis,moon2014ask,dutta2021msfl,combifl} in addition to having low efficiency, entail a considerable amount of space which sometimes makes it practically infeasible for large-sized programs.
\par \textcolor{black}{Statistical Models have been immensely manipulated in the past for accomplishing the objective of better fault localization. Crosstab proposed by Wong et al.\cite{wong2008crosstab} and FTFL proposed by Dutta et al.\cite{ftfl} have highlighted the dominance of statistical models in producing effective fault localizers. Therefore, in this article, we extensively employ conditional probability statistics to evaluate the suspicious score of the program statements. Conditional Probability statistic\cite{cpfl} portrays it's significance by highlighting the correlation between the program spectra and execution results. Further, we make use of the grouping method to refine the rank of the buggy statement. It is also a relatively simpler statistical model as compared to Crosstab\cite{wong2008crosstab} and FTFL\cite{ftfl}. Hence, we name our proposed method as Conditional Probability and Grouping based Fault Localization technique(hereafter referred as CGFL). }
\par Rest of the paper is organized as follows. In Section \ref{rw}, we present a survey of related literature. We discuss our proposed approach in Section \ref{pw}. In Section \ref{es}, we elaborate upon the experimental result followed by the comparison of our proposed technique with related work in Section \ref{crw}. Finally, we conclude in Section \ref{conc}. 

\section{Related Work}\label{rw}

\par Weiser introduced the concept of static slicing\cite{weiser1984program}. Static slicing\cite{weiser1984program} was based on the design that if a test case fails in response to a variable attaining the wrong value at a statement, then the slice associated with the variable-statement pair would be held accountable for the resulting defect. Later, Korel et al. \cite{korel1988dynamic}, proposed dynamic slicing which eliminated the drawback of static slicing where all the executable statements which could potentially be affected by the value of a variable at a statement were included. Hence, in dynamic slicing \cite{korel1988dynamic} only those statements were included which actually influenced the value of a variable at a statement. But in some cases it may so happen that the slice returned is very large(sometimes even the size of the entire program) which defeats the overall purpose of using this technique.
\par Next, Spectrum Based Fault Localization(SBFL) methods were proposed which gradually became popular. The SBFL techniques have low computation costs  as these methods generally require a mathematical expression for the computation of suspicious score for each executable program statement. They also utilize various statistics to produce a mathematical model which could determine the suspicious score of  program statements. Tarantula is one of the most prominent SBFL techniques proposed by Jones et al. \cite{jones2005empirical}. It outperformed set union\cite{setunionintersection}, set intersection\cite{setunionintersection}, nearest neighbour\cite{setunionintersection}, and cause transition\cite{causetransition} which were some of the pre-existing effective FL techniques at that time. But eventually, it was realized that Tarantula was incapable of perfectly exploiting the relevant information carried by the successful and failed test case execution results. To overcome the shortcomings of Tarantula\cite{jones2005empirical}, several other SBFL approaches have been proposed. These included Ample\cite{wong2016survey}, Jaccard\cite{wong2010family}, Ochiai\cite{naish2011model}, Barinel and DStar(D$^*$)\cite{wong2013dstar}, etc \cite{??}. Ample\cite{wong2016survey} and Jaccard\cite{wong2010family} use failed as well as passed test cases into consideration while calculating the suspicious score of statements. Ochiai\cite{naish2011model}, which is basically obtained from the domain of molecular biology, puts a greater emphasis on the failed test cases which were responsible for not executing the statement. Wong et al.\cite{wong2013dstar} developed DStar(D$^*$) which showed better performance as compared to all the pre-existing techniques and thus became the state-of-the-art technique. The value of `*' in D$^*$ was chosen as 2 as it exhibited the best results. But since the DStar technique took a small number of parameters into consideration while calculating the suspicious score, many program statements were assigned the same rank.
\par Machine Learning (ML) played a key role in the further advancement of FL techniques. The highly adaptive nature of machine learning algorithms helped produce robust models which were found to be very effective. ML techniques generally, learn the program spectra and execution results and generate suspiciousness scores for program elements based on their learning. Back Propagation Neural Network(BPNN) technique proposed by Wong et al.\cite{wong2009bp} is one of the most prevalent machine learning-based FL techniques. It has easy implementation due to its very fundamental structure. But it was noticed that BPNN \cite{wong2009bp} suffered from problems of local minima\cite{tan2013data}. and paralysis\cite{wasserman1993advanced}. Hence to resolve this issue, Wong et al.\cite{wong2011effective} proposed the Radial Basis Function Neural Network(RBFNN) technique which had radial basis function instead of sigmoid function as its transfer function. But, BPNN\cite{wong2009bp} and RBFNN\cite{wong2011effective} were not found effective in handling large complex functions because of their shallow architecture. So, Zheng et al.\cite{zheng2016fault} develop Deep Neural Network(DNN)\cite{zheng2016fault} model which made working with complex functions relatively simpler. Later, Dutta et al.\cite{dutta2019hierarchically} further refined the technique by proposing a hierarchical approach of FL using DNN where they initially investigated the functions for containing a fault and then the statements. But these neural network-based approaches desired training of large number of parameters which demonstrated a very complex model for solving the fault localization problem.
\par In last four to five years, several deep learning-based FL techniques such as DeepFL\cite{li2019deepfl} and DeepRL4FL\cite{li2021fault} have been proposed. These FL techniques incorporates several information obtained from different traditional FL
methods, for example the suspiciousness scores calculated from the SBFL and MBFL techniques, text similarity, static code metrics etc. These techniques utilizes the learning capability of neural networks to train the classification models for accurately localizing the faulty program entities. DeepFL\cite{li2019deepfl} make use of the synthetically designed features resulting in underutilization of program spectra, rather than considering the contextual information between program entities. Also, this technique uses several complex features e.g. spectrum-based suspiciousness and complexity-based fault proneness which results in higher overhead for obtaining the required information.
\par In the recent past, researchers have laid emphasis on the usage of mutation analysis for the purpose of fault localization. Papadakis et al. \cite{papadakis2015metallaxis} introduced the mutation-based FL technique known as Metallaxis-FL\cite{papadakis2015metallaxis}, where they generated mutants of the program in such a manner that if a generated mutant was killed by the failed test case then that particular mutant would provide an excellent indication of the faulty location in the program. Later, Moon et al.\cite{moon2014ask} developed a technique called MUSE\cite{moon2014ask}, which primarily identified faulty statements by making use of the different characteristics in two broad groups of mutants. The first group consisted of mutants which were generated by mutating the faulty statement and the second group comprised of mutants generated by mutating the non-faulty statements of the program. Also, Dutta et al. proposed MBFL techniques such as MSFL\cite{dutta2021msfl} and Combi-FL\cite{combifl} which suggested significant improvement to the pre-existing mutation-based techniques. Although these mutation-based fault localization techniques proved to be effective, they required a huge computation cost due to the large number of mutants generated for large programs. Not only that, these heavy computations also lowered the efficiency of the technique.  

\section{Proposed Work: CGFL}\label{pw}

 \par The necessity of fault localization arises as soon as a test case execution failure is reported by the program. It signals that the program is faulty. The very first step towards addressing the problem of fault localization is to execute a large number of test cases on the faulty program. Subsequently, the program spectra information is extracted. The fault localization technique utilizes the program spectra information obtained to generate a ranked list of suspicious scores. Each element of this list indicates the suspicious score of each program entity. For simplicity, we have chosen executable statements as the program entity. In this section, we first describe our proposed technique CGFL.  Subsequently, we present an illustration of our proposed approach using an example program. Table \ref{tab:notations} characterizes the notations used in this paper. We denote the executable program statement with the symbol $\zeta$.

\begin{table}[htp]
    \centering
    \caption{Notations Used in this paper}
    \label{tab:notations}
    \begin{tabular}{|c|c|}
         \hline
         $\tau$ & Total number of test cases\\
         \hline
         $\tau_f$ & Total number of failed test cases\\
         \hline
         $\tau_s$ & Total number of successful test cases\\
         \hline
         $\tau_c(\zeta)$ & Number of test cases covering $\zeta$\\
         \hline
         $\tau_{cf}(\zeta)$ & Number of failed test cases covering $\zeta$\\
         \hline
         $\tau_{cs}(\zeta)$ & Number of successful test cases covering $\zeta$\\
         \hline
         $\tau_u(\zeta)$ & Number of test cases not covering $\zeta$\\
         \hline
         $\tau_{uf}(\zeta)$ & Number of failed test cases not covering $\zeta$\\
         \hline
         $\tau_{us}(\zeta)$ & Number of successful test cases not covering $\zeta$\\
         \hline
    \end{tabular}
\end{table}

\subsection{Overview}

The internal linkage between the test case execution result and
program spectrum is the fundamental basis for constructing any  statement suspiciousness calculating formula. It is always important as well as essential to capture all possible correlations between statement coverage (covered/uncovered) and test result (pass/fail).  Conditional probability in statistics effectively captures the likelihood of the occurrence an event based on the the occurrence of a previous outcome or event.
The conditional probabilities provide extremely useful information, even when limited information is provided. In this paper, we design four conditional probability
models to capture the  association between the program spectrum and test execution results. Using these four models and the concepts of the occurrence
of low-probability events in the information theory, we define a new probability based fault localization technique.

We construct a probabilistic model to depict the association  of test case execution result(i.e., pass or fail) with the execution of program statements by that particular test case and vice versa as well. A  statistic(denoted by $\psi$) is designed for each dependency relationship. We define 4-$\psi$ statistics which are capable enough to consider all relevant dependency scenarios. In the definition of 4-$\psi$ statistics, $C$ denotes that the statement was executed, $U$ denotes that the statement was not executed, $F$ denotes that the test case was failed and $S$ denotes that the test case was successful. When the actual test case output is different from the expected output then the test  case is considered as fail (unsuccessful) otherwise pass (successful).

\begin{enumerate}
    \item Statistic-1($\psi_{fc}$): This statistic computes the probability of the test case failure when it is known that the statement has been executed by the test case. Mathematically, it can be represented as:
    \begin{align*}
        \psi_{fc}(\zeta)\,=\,P(F|C)\,=\frac{\tau_{cf}(\zeta)}{\tau_{cf}(\zeta)+\tau_{cs}(\zeta)} 
    \end{align*}
    where $\tau_{cf}(\zeta)+ \tau_{cs}(\zeta) \neq 0$. If $\tau_{cf}(\zeta)+ \tau_{cs}(\zeta) = 0$, it exhibits that the statement $\zeta$ was not covered by any of the test cases and this situation is likely to provide no information for fault localization.\newline
    \item Statistic-2($\psi_{cf}$): This model determines the probability of a test case to execute the statement $\zeta$ when it is noted that the execution of the test case resulted in a failure. Mathematically, it can be represented as:
    \begin{align*}
        \psi_{cf}(\zeta)\,=\,P(C|F)\,=\frac{\tau_{cf}(\zeta)}{\tau_{cf}(\zeta)+\tau_{uf}(\zeta)} 
    \end{align*}
    where $\tau_{cf}(\zeta) + \tau_{uf}(\zeta) \neq 0$. If $\tau_{cf}(\zeta)+ \tau_{uf}(\zeta) = 0$, it indicates that all of the test cases have led to execution failure and such a state of condition is meaningless from that fault detection point of view.\newline
    \item Statistic-3($\psi_{cs}$): This statistic is responsible for evaluating the probability of a test case to execute the statement $\zeta$ when it is well known that the execution of the test case resulted in a success. Mathematically, it can be represented as:
    \begin{align*}
        \psi_{cs}(\zeta)\,=\,P(C|S)\,=\frac{\tau_{cs}(\zeta)}{\tau_{cs}(\zeta)+\tau_{us}(\zeta)} 
    \end{align*}
    where $\tau_{cs}(\zeta) + \tau_{us}(\zeta) \neq 0$. If $\tau_{cs}(\zeta) + \tau_{us}(\zeta) = 0$, it manifests that all of the test cases have lead to execution success and such a situation does not contribute any relevant information to the fault localization technique.\newline 
    \item Statistic-4($\psi_{su}$): This model quantifies the probability of a test case having a successful execution result when it has already been recognized that the test case did not execute the statement $\zeta$. Mathematically, it can be represented as:
    \begin{align*}
        \psi_{su}(\zeta)\,=\,P(S|U)\,=\frac{\tau_{us}(\zeta)}{\tau_{uf}(\zeta)+\tau_{us}(\zeta)} 
    \end{align*}
    where $\tau_{uf}(\zeta) + \tau_{us}(\zeta) \neq 0$. If $\tau_{uf}(\zeta) + \tau_{us}(\zeta) = 0$, it reveals that the statement $\zeta$ was covered by all the test cases, thereby leaving little scope for the fault localization technique to extract some useful information.\newline
\end{enumerate}

\par These 4-$\psi$ statistic models would be supportive in constructing the fault localizer (denoted by $\chi$). The fault localizer $\chi$ would generate the suspicious score for each of the program statements. The fault localizer $\chi$ is defined as follows:

\[
    \chi(\zeta) = 
    \begin{cases} 
        -\infty, & \text{if } \psi_{fc}=0 \text{ or } \psi_{su}=0\\
        \psi_{fc}+\psi_{cf}+\psi_{su}, & \text{if } \psi_{fc}\neq0 \text{ and } \psi_{su}\neq0\\
    \end{cases}
\]

 The suspicious score of the program statements with  $\psi_{fc}$ or $\psi_{su}$ values as zero is considered to be $-\infty$. This is because $\psi_{fc}=0$ points to a situation where the probability of a test case to fail is always zero given that it has executed a particular statement. Then, such a statement is least likely to be faulty and hence is assigned with the least priority. Similarly, when $\psi_{su}=0$, it represents a situation where the probability of a test case to pass is always zero given that it does not execute a particular statement. Similarly, in this situation also, that particular statement holds the least probability to be faulty, and therefore, its suspicious score value is assigned as $-\infty$. The fault localization technique  $\chi$ is also termed as CPFL which stands for Conditional Probability based Fault Localization. The fault localizer $\chi$ generates a listed of suspicious scores of program statements. The statements can be ranked from higher to lower based on their degree of suspiciousness. However, we have further extended the FL technique $\chi$ by incorporating the grouping strategy and then termed it as CGFL where represents the combination of Conditional probability and Grouping strategy for Fault Localization. The grouping strategy improves the ranking strategy of the CPFL. We demonstrate the improvement obtained in CGFL over CPFL in the experimental section \ref{es}.

\par There is a popular intuition in fault localization that if a program statement has been covered by a large number of failed test cases then there is a high possibility that the statement may be buggy. Whereas, if the statement has been covered by a small number of failed test cases, the chance of the statement being buggy turns out to be low. Therefore, we prioritize the statements which have been executed by a large number of failed test cases for examination using the grouping strategy \cite{debroy2010grouping}. The program statements are grouped subsequently, such that each group consists of all the statements that are executed by a particular number of failed test cases. For example, if a program has $n$ number of failed test cases. Then, at most $n+1$ groups $g_0$, $g_1$, $g_2$, ..., $g_n$ would be formed. 
\par The groups are created in such a manner that group $g_0$ contains all the statements that have not been executed by any of the failed test cases, group $g_1$ is a collection of all the statements that have been executed by only one failed test case and similarly group $g_n$ consists of all the statements that have been executed by all(i.e. $n$) failed test cases. Also, the grouping of statements is purely based on the cardinality of the failed test cases which have executed the particular statement. Let us say, we have two statements $\zeta_1$ and $\zeta_2$ such that they were executed by failed test cases \{$t_1$,$t_3$,$t_7$\} and \{$t_1$,$t_2$,$t_9$\} respectively. Even though both the statements are executed by different sets of failed test cases yet they both would be assigned to the same group $g_3$(as they were executed by three failed test cases).
The statements residing in the same group are then arranged in the descending order of their suspicious score. The suspicious score was computed by the fault localizer $\chi$. 
\begin{figure}
	\centering
	\includegraphics[width=1.0\linewidth,height=0.45\textheight]{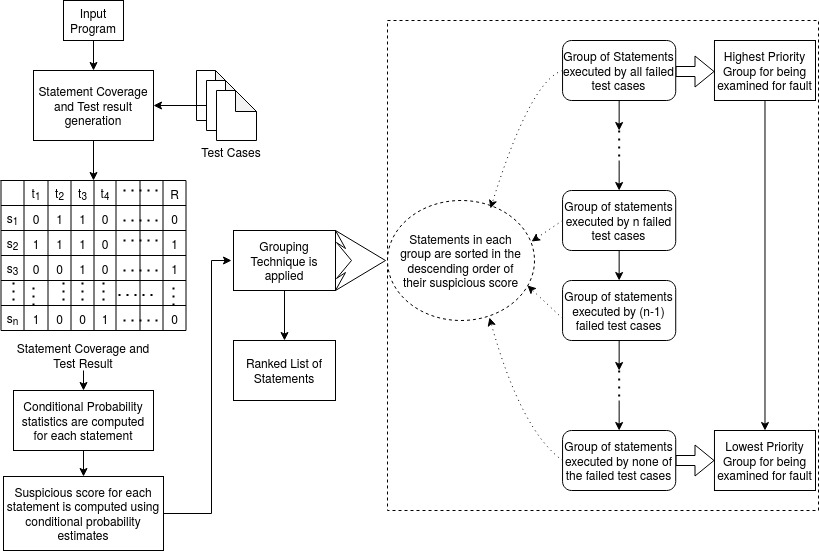}
	\caption{Block diagram of proposed CGFL technique}
	\label{fig:cgfl-page-3}
\end{figure}
\par Figure \ref{fig:cgfl-page-3} represents the block diagram of proposed CGFL technique. The walk though of this diagram and the workflow of the complete CGFL approach is explained through the  following set of steps.
\begin{enumerate}[leftmargin=34pt, labelsep = 0.5em]
	\item[Step 1:] (Computation of Program Spectra and Test Results)\newline
	We first supply the input program and run it on the given test suite to generate the statement coverage information and test execution results.
    \item[Step 2:] (Calculation of suspicious score)\newline
    Subsequently, statement coverage data and test results are supplied to the fault localizer $\chi$. Using the constructed 4-$\psi$ statistics, $\chi$ evaluates the suspicious score $\chi(\zeta)$ for each  executable program statement $\zeta$.
    \item[Step 3:] (Assignment of statements to groups)\newline
    further, each statement $\zeta$ is assigned to the group, which is responsible for holding all the statements that have been covered by a particular number of failed test cases.
    \item[Step 4:] (Sorting of Groups)\newline
    The groups are sorted in such a manner that the one which is a collection of statements executed by larger number of failed test cases get a higher priority in comparison to one which holds statements executed by smaller number of failed test cases. For example, $g_7$ will have more preference for fault localization than $g_5$ or $g_6$.
    \item[Step 5:] (Sorting of statements within each group)\newline
    The statements within each group are sorted in descending order of their suspicious score.
\end{enumerate}

This completes the description of the proposed CGFL technique.

\subsection{Example}
In this section,  we illustrate the working of our proposed CGFL technique using an example program. Table \ref{tab:table1} presents a simple code snippet along with the complete statement coverage and test case execution result for the example program. Each row (starting from $4^{th}$ row) of Table \ref{tab:table1} represents an executable program statement and its corresponding coverage information for all test cases. In the Table, `1' denotes that the test case executed the statement while `0' denotes that the test case did not execute the statement. The last row of Table \ref{tab:table1} manifests the execution result for each test case. Here, F and P denote that the corresponding test case has failed and passed respectively.

\begin{table}[ht]
\centering
\caption{Program illustrating CGFL technique}
\begin{tabular}{|c|c|c|c|c|c|c|c|c|c|c|c|}
\hline
Program & \multicolumn{11}{c|}{Test Cases}  \\ \hline
 & 1 & 2 & 2 & 1 & 2 & 1 & 2 & 2 & 3 & 3 & 2 \\ \cline{2-12}
 & 2 & 2 & 1 & 3 & 3 & 2 & 1 & 1 & 1 & 2 & 2 \\ \cline{2-12}
\multirow{-3}{*}{\begin{tabular}[c]{@{}c@{}}int find\_mid(int\\  p, int q, \\ int r)\{\end{tabular}} & 3 & 2 & 3 & 2 & 1 & 1 & 2 & 1 & 2 & 1 & 1  \\ \hline \hline
int mid;                & 1  & 1  & 1  & 1  & 1  & 1  & 1  & 1  & 1  & 1  & 1  \\ \hline
mid=r;                  & 1  & 1  & 1  & 1  & 1  & 1  & 1  & 1  & 1  & 1  & 1  \\ \hline
if (q \textless r)      & 1  & 1  & 1  & 1  & 1  & 1  & 1  & 1  & 1  & 1  & 1  \\ \hline
if (p \textgreater q) //bug   & 1  & 0  & 1  & 0  & 0  & 0  & 1  & 0  & 1  & 0  & 0  \\ \hline
mid = q;           & 0  & 0  & 1  & 0  & 0  & 0  & 1  & 0  & 1  & 0  & 0  \\ \hline
else if (p \textless r) & 1  & 0  & 0  & 0  & 0  & 0  & 0  & 0  & 0  & 0  & 0  \\ \hline
mid = p;                & 1  & 0  & 0  & 0  & 0  & 0  & 0  & 0  & 0  & 0  & 0  \\ \hline
else if (p \textgreater q) & 0  & 1  & 0  & 1  & 1  & 1  & 0  & 1  & 0  & 1  & 1  \\ \hline
mid = q;                & 0  & 0  & 0  & 0  & 0  & 0  & 0  & 1  & 0  & 1  & 0  \\ \hline
else if (p \textgreater r) & 0  & 1  & 0  & 1  & 1  & 1  & 0  & 0  & 0  & 0  & 1  \\ \hline
mid = p;                & 0  & 0  & 0  & 0  & 1  & 0  & 0  & 0  & 0  & 0  & 1  \\ \hline
return mid;             & 1  & 1  & 1  & 1  & 1  & 1  & 1  & 1  & 1  & 1  & 1  \\ \hline
\} & 1  & 1  & 1  & 1  & 1  & 1  & 1  & 1  & 1  & 1  & 1  \\ \hline \hline
Fail/Pass              & F  & P  & F  & P  & P  & P  & F  & P  & F  & P  & P  \\ \hline
\end{tabular}
\label{tab:table1}
\end{table}

\par After obtaining the complete statement coverage information, the subsequent step is to determine the values of each statistic for each program statement $\zeta$. Values of the different parameters required to compute the suspicious score of each statement are shown in Table \ref{tab:table2}.

\begin{table*}[ht]
    \centering
    \caption{Parameter Values for the example Program}
    	\scalebox{0.7}{\renewcommand{\arraystretch}{1.5}\begin{tabular}{|c|c|c|c|c|c|c|c|c|c|c|c|c|c|}
        \hline
        Parameters & $\zeta_1$ & $\zeta_2$ & $\zeta_3$ & $\zeta_4$ & $\zeta_5$ & $\zeta_6$ & $\zeta_7$ & $\zeta_8$ & $\zeta_9$ & $\zeta_{10}$ & $\zeta_{11}$ & $\zeta_{12}$ & $\zeta_{13}$ \\ \hline
        $\tau_{cf}(\zeta)$ & 4 & 4 & 4 & 4 & 3 & 1 & 1 & 0 & 0 & 0 & 0 & 4 & 4 \\ \hline
        $\tau_{cs}(\zeta)$ & 7 & 7 & 7 & 0 & 0 & 0 & 0 & 7 & 2 & 5 & 2 & 7 & 7 \\ \hline
        $\tau_{uf}(\zeta)$ & 0 & 0 & 0 & 0 & 1 & 3 & 3 & 4 & 4 & 4 & 4 & 0 & 0 \\ \hline
        $\tau_{us}(\zeta)$ & 0 & 0 & 0 & 7 & 7 & 7 & 7 & 0 & 5 & 2 & 5 & 0 & 0 \\ \hline
        $\psi_{fc}(\zeta)$ & 0.36 & 0.36 & 0.36 & 1 & 1 & 1 & 1 & 0 & 0 & 0 & 0 & 0.36 & 0.36\\ \hline
        $\psi_{cf}(\zeta)$ & 1 & 1 & 1 & 1 & 0.75 & 0.25 & 0.25 & 0 & 0 & 0 & 0 & 1 & 1 \\ \hline
        $\psi_{cs}(\zeta)$ & 1 & 1 & 1 & 0 & 0 & 0 & 0 & 1 & 0.29 & 0.71 & 0.29 & 1 & 1 \\ \hline
        $\psi_{su}(\zeta)$ & not def. & not def. & not def. & 1 & 0.88 & 0.7 & 0.7 & 0 & 0.56 & 0.33 & 0.56 & not def. & not def. \\ \hline
    \end{tabular}}
    \label{tab:table2}
\end{table*}

\par Subsequently, the suspicious score for each executable program statement is calculated with the aid of fault localizer $\chi$. Table \ref{tab:table3} shows the computed suspicious scores for the program statements.

\begin{table*}[ht]
    \centering
    \caption{Suspicious score of program statements}
    \scalebox{0.72}{\renewcommand{\arraystretch}{1.3}\begin{tabular}{|c|c|c|c|c|c|c|c|c|c|c|c|c|c|}
        \hline
        Statements & $\zeta_1$ & $\zeta_2$ & $\zeta_3$ & $\zeta_4$ & $\zeta_5$ & $\zeta_6$ & $\zeta_7$ & $\zeta_8$ & $\zeta_9$ & $\zeta_{10}$ & $\zeta_{11}$ & $\zeta_{12}$ & $\zeta_{13}$ \\ \hline
        $\chi(\zeta)$ & $-\infty$ & $-\infty$ & $-\infty$ & 3.0 & 2.625 & 1.95 & 1.95 & $-\infty$ & $-\infty$ & $-\infty$ & $-\infty$ & $-\infty$ & $-\infty$ \\ \hline
    \end{tabular}}
    \label{tab:table3}
\end{table*}
\begin{table}
    \centering
    \caption{Statements Assigned to each group}
    \begin{tabular}{|c|c|}
        \hline
        Group no. & Statements arranged in descending order of $\chi(\zeta)$ value \\ \hline
        $g_4$ & $ \chi(\zeta_4) > \chi(\zeta_1) = \chi(\zeta_2) = \chi(\zeta_3) = \chi(\zeta_{12}) = \chi(\zeta_{13}) $ \\ \hline
        $g_3$ & $ \chi(\zeta_5) $ \\ \hline
        $g_2$ & None \\ \hline
        $g_1$ & $ \chi(\zeta_6) = \chi(\zeta_7) $ \\ \hline
        $g_0$ & $ \chi(\zeta_8) = \chi(\zeta_9) = \chi(\zeta_{10}) = \chi(\zeta_{11}) $ \\ \hline
    \end{tabular}
    \label{tab:table4}
\end{table}  
\par Now, the grouping approach is implemented for refining the rank of the buggy statement. There are 4 failed test cases and 7 successful test cases in the example program. So, there would be 5 groups formed($g_0$, $g_1$, $g_2$, $g_3$, $g_4$). The statements assigned to each group are represented in Table \ref{tab:table4}.

\par According to obtained results for the example, it is found that the statement $\zeta_4$ receives highest priority for being examined for a bug. This is in resemblance to the veracity that statement $\zeta_4$ was faulty. Therefore, this example explains the complete methodology adopted to implement the CGFL technique.

\section{Experimental Results}\label{es}
\par In this section, we first present the experimental setup and the data set used for experimentation. We then listed the evaluation metrics used to determine the performance of the proposed CGFL approach. Subsequently, the experimental results are discussed. Finally, we complete this section by highlighting the threats to the validity of the obtained results.

\subsection{Setup}
A 64-bit Ubuntu machine having specifications of 15.8 GB RAM and Intel (R) Core(TM)-i7 processor is employed for carrying out all the experiments. One set of input programs i.e., Siemens suite\cite{siemens} are written in ANSI-C format. On the other hand, Defects4j\cite{Defects4J}, the other program suite contains Java programs. The statement coverage matrix and test case execution results are generated for each faulty program using GCOV \cite{gcov_man_page} tool for the C-format programs. GCOV is a utility tool that comes as a product of the GNU Compiler Collection suite. It is primarily used for code coverage analysis and statement-by-statement profiling of C-programs. For Defects4j\cite{Defects4J}, we have used open-source available coverage results and other required resources in our experiment from \cite{Defects4Jmut}.

\begin{table}
	\centering
	\caption { Program characteristics}
	\scalebox{0.9}
	{\renewcommand{\arraystretch}{1.2 }\begin{tabular}{|c|c|c|c|c|c|}
	        \hline 
	        S. & Program & LOC & No. of & No. of & No. of \\ 
	        No. & Name &  & Exec. Lines & Flty Vers. & Test Cases \\ \hline
	        1.1 & Print\_Tokens & 565 & 195 & 7 & 4130 \\ \hline
	        1.2 & Print\_Tokens2 & 510 & 200 & 10 & 4115 \\ \hline
	        1.3 & Schedule & 412 & 152 & 9 & 2650 \\ \hline
	        1.4 & Schedule2 & 307 & 128 & 10 & 2710 \\ \hline
	        1.5 & Replace & 521 & 244 & 32 & 5542 \\ \hline
	        1.6 & Tcas & 173 & 65 & 41 & 1608 \\ \hline
	        1.7 & Tot\_info & 406 & 122 & 23 & 1052 \\ \hline
	        2.1 & Lang & 39.8K & 30.2K & 64 & 2245 \\ \hline
 			2.2 & Math & 45K & 19K & 104 & 3602 \\ \hline
 			2.3 & Mockito & 27.8K & 19.8K & 38 & 5205 \\ \hline
 			2.4 & Time & 56.2K & 40.1K & 26 & 4130 \\ \hline
	\end{tabular}  }
	\label{programsummary}
\end{table} 
\subsection{Used Data set}
	\par In order to analyze the effectiveness of our proposed method CGFL, we considered two different program suites for experimentation. The first program suite is \textit{Siemens suite} which comprises of total seven programs. The Siemens suite programs are extracted from \textit{SIR repository} \cite{siemens}. The second program suite is Defects4j\cite{Defects4J}. We took four programs from \textit{Defects4j} repository into consideration. The test suites and buggy versions are already present in these benchmark suites. 
	\par Table \ref{programsummary} presents a tabular representation of program characteristics. Colu- mns 3-6 indicate the  total number of lines (LOC) present in the program, total executable lines of code present, number of faulty versions available, and number of test cases present in the test suite of that program.
	\par Siemens suite has been widely used in the past as a benchmark for evaluating the effectiveness of various fault localization techniques \cite{wong2008crosstab,wong2013dstar,jones2002visualization}.  \textit{Print\_Tokens} and \textit{Print\_Tokens2} programs are prominent lexical analyzers.  Siemens programs \textit{Schedule2} and \textit{Schedule} perform the task of priority scheduling. The \textit{Replace} program is popularly used for pattern matching and substitution. \textit{Tot\_info} finds its application in computing several statistics of input data.  \textit{Tcas} stands for Traffic Collision Avoidance system and these programs are designed for minimizing the chances of mid-air collision between aircrafts. 
\par The remaining four different Java programs have been taken from Defects4j \cite{Defects4J} repository. \textit{Lang} is a  Java utility class that helps to design and model the Java language. \textit{Math} is a lightweight and self-contained  library of mathematics and statistics components. \textit{Mockito} is used to test the unit Java classes. \textit{Time} (aka Joda-Time) is the de facto standard time and date library for Java programs.
	
	\par The Siemens suite consists of a total of 132 faulty versions from seven different programs. However, we have considered only 116 faulty versions and omitted 16 versions, due to one or more of the following scenarios:
\begin{itemize}
	\item A modification was made in a non-executable patch of the program. 
	\item No semantic difference exists between the original and faulty programs except the header files.
	\item A test case execution resulted in a segmentation fault.
	\item None of the test cases failed for the version.
\end{itemize}

\par Now, for determining if a test case is successful or it has failed, the following steps were implemented.
\begin{enumerate}
	\item All the test cases present in the test suite were executed with the faultless program(i.e. original program) and the output generated for each test case was stored in separate files.
	\item  The same test cases were executed using a faulty version of the program and the generated output was saved in a similar fashion.
	\item  Finally, a comparison was made to find out if the outputs generated for the fault-free and buggy version of the program are exactly the same or not. If the outputs were exactly identical then it is considered that the test case is successful; else, the test case is deemed to be failed for that particular faulty version.
\end{enumerate}

\subsection{Evaluation Metric}
\par We make use of four different metrics for analyzing the effectiveness of our proposed CGFL method. The following are the metrics:
\par \textbf{EXAM\_Score}: This metric is popularly used for determining the effectiveness of a fault localization technique\cite{ftfl,setunionintersection}. Equation \ref{escore} presents the formula for computing EXAM\_Score.
\begin{equation}\label{escore}
    EXAM\_Score = \frac{|S_{examined}|}{|S_{total}|}*100
\end{equation}
where $|S_{examined}|$ shows the number of statements examined to localize the fault and $|S_{total}|$ denotes the total statements present in the program. Let us assume, on a program P, we used two fault localization techniques, say, $T_a$ and $T_b$, and  obtained EXAM\_Scores are $ES_a$ and $ES_b$ respectively. If $ES_a$ is less than $ES_b$, then technique $T_a$ is more effective than $T_b$.
\par \textbf{Top-N\%:} This metric shows the percentage of faulty versions which are correctly localized by examining at most N\% of executable program statements\cite{li2019deepfl}. For this study, we consider the N values as 1 and 5. It helps to discover the FL technique that localizes more faulty versions by analyzing up to 1\% or 5\% of program statements.
\par \textbf{Relative Improvement (RImp):} It shows the comparative improvement obtained using CGFL technology over existing FL methods. It is calculated using Equation \ref{iai}.
\begin{equation}\label{iai}
    RImp_{a,b}=\frac{\#\ statements\ examined\ by\ Technique_a}{\#\ statements\ examined\ by\ Technique_b}*100
\end{equation}
\par In this equation, Technique$_a$ represents our proposed CGFL method and Technique$_b$ can be any of the existing FL technique such as DStar\cite{wong2013dstar}, Tarantula \cite{jones2005empirical} etc. with which proposed CGFL is compared. If the number of statements examined by Technique$_a$  is lesser than Technique$_b$ then the value of RImp$_{a,b}$ is lesser than 100\% otherwise it is more than 100\%.
\par \textbf{Average Improvement:}  \textcolor{black}{ This metric indicates the average improvement realized on using an FL technique over another FL technique\cite{combifl}. It is computed using Equation~\ref{eq3}.
\begin{equation}\label{eq3}
IA_{a,b}=\frac{Avg.ES_b-Avg.ES_a}{Avg.ES_a}*100
\end{equation}
Where, $IA_{a,b}$ exhibits the improvement achieved using $T_a$ over $T_b$. $Avg.ES_a$ and $Avg.ES_b$ indicate the average $ES$ obtained by $T_a$ and $T_b$ respectively. Hence, the lesser the average $EXAM\_Score$ better the technique. }
\subsection{Results}
\begin{figure}
    \centering
    \includegraphics[width=0.85\textwidth]{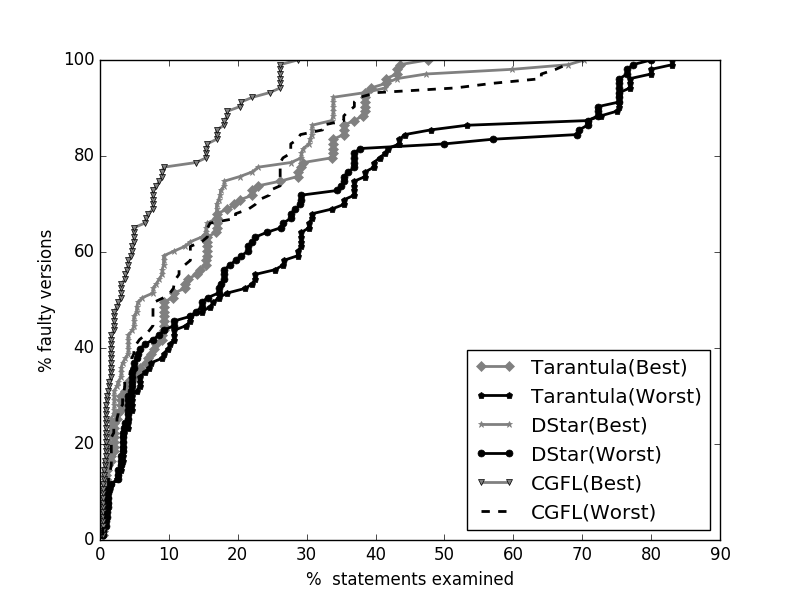}
		\caption{Effectiveness of CGFL with DStar and Tarantula for Siemens  suite }
    \label{cgf_siemens_dt}
\end{figure}
\par In this section, we discuss the obtained comparative  results for the proposed method CGFL with eight existing FL techniques. We consider four techniques from spectrum based family viz., Tarantula\cite{jones2002visualization}, DStar\cite{wong2013dstar}, Ochiai\cite{naish2011model}, and Crosstab\cite{wong2008crosstab} and remaining four methods from neural network models which includes BPNN\cite{wong2009bp}, DNN\cite{zheng2016fault}, RBFNN\cite{wong2011effective} and CNN\cite{zhang2019cnn}. Tarantula\cite{jones2002visualization} is a classic SBFL technique. Crosstab\cite{wong2008crosstab} is a statistics-based SBFL method that uses the Chi-square test to decide the dependency of test case result (success/failure) on the invocation of a specific program statement. Ochiai\cite{naish2011model} is another prominent SBFL technique. DStar\cite{wong2013dstar} is known as one of the state-of-the-art SBFL methods. Among the neural network-based FL models, BPNN\cite{wong2009bp} is the most simple and easiest to implement. It was the first model used for FL. RBFNN\cite{wong2011effective} deals with the problems of paralysis\cite{wasserman1993advanced} and local minima\cite{tan2013data}. DNN\cite{zheng2016fault} and CNN\cite{zhang2019cnn} are the two most robust and effective models for fault localization.
\par Some of the FL techniques may assign identical suspicious scores to multiple program elements. This leads to two different types of effectiveness for that FL method viz., the \textit{best} effectiveness and the \textit{worst} effectiveness. When the fault localizer points to the buggy line first for examination, among all the statements holding the same suspiciousness values then it is termed as its \textit{best} effectiveness. On the other hand, when the fault localizer points to the buggy line at last for examination, among all the statements holding the same suspiciousness values then it is termed as its \textit{worst} effectiveness.
\begin{figure}
    \centering
    \includegraphics[width=0.85\textwidth]{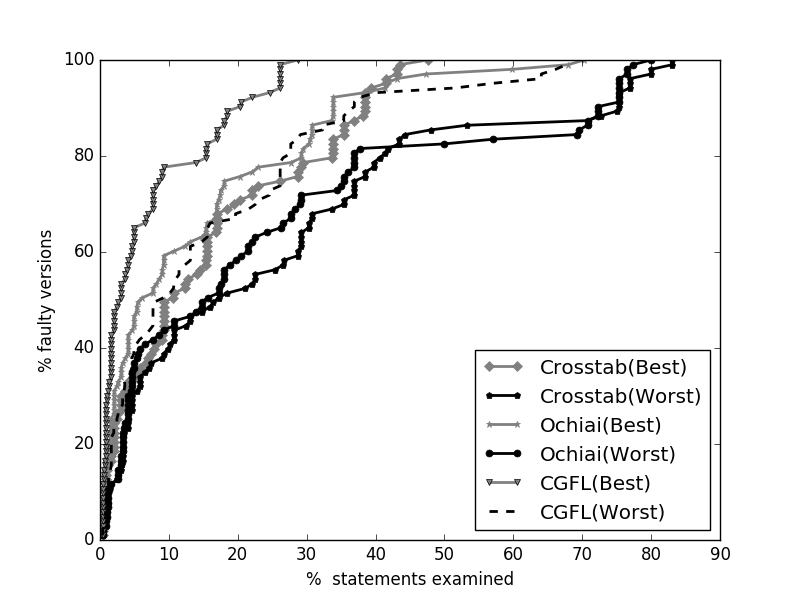}
		\caption{Effectiveness of CGFL with Crosstab and Ochiai for Siemens  suite}
	\label{cgf_siemens_co}
\end{figure}
\par
\begin{figure}
    \centering
    \includegraphics[width=0.85\textwidth]{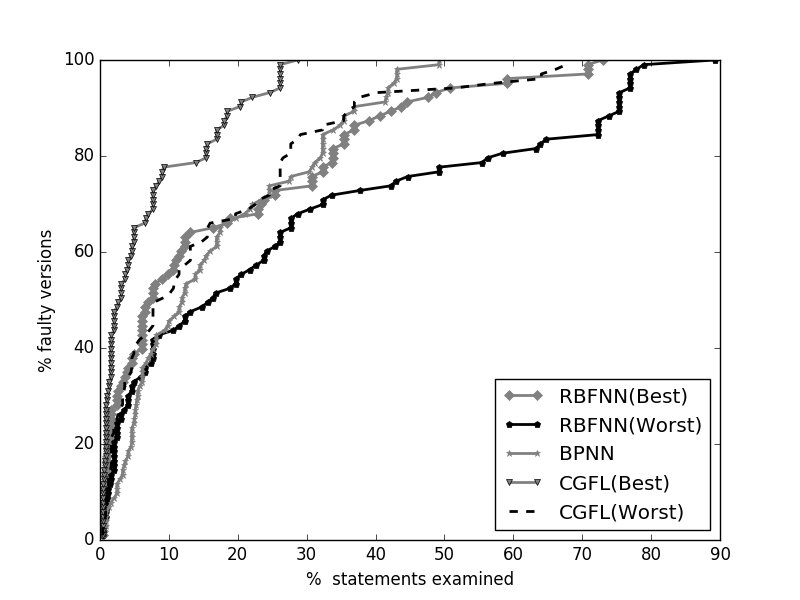}
		\caption{Effectiveness of CGFL with RBFNN and BPNN for Siemens  suite }
		\label{cgf_siemens_rb}
\end{figure}

\begin{figure}
    \centering
    \includegraphics[width=0.85\textwidth]{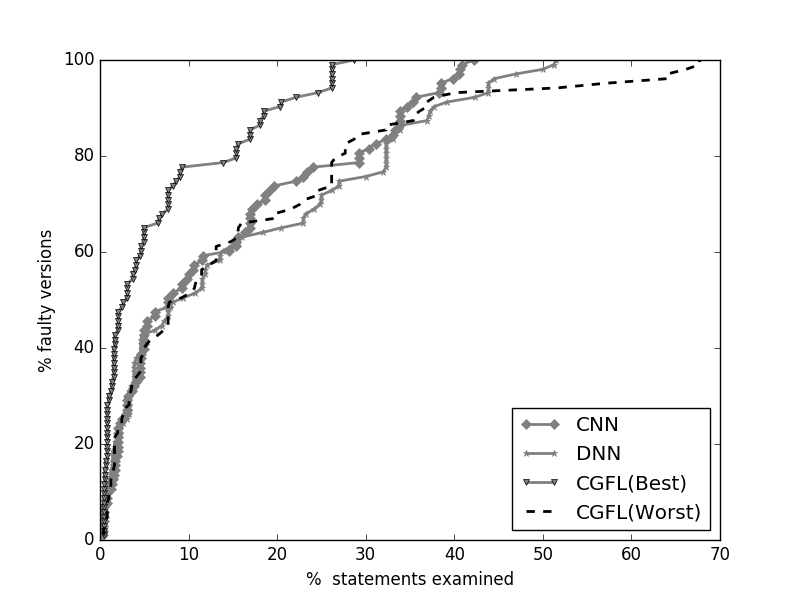}
		\caption{Effectiveness of CGFL with DNN and CNN  for Siemens  suite }
		\label{cgf_siemens_dc}
\end{figure}

\begin{figure}
    \centering
    \includegraphics[width=0.85\textwidth]{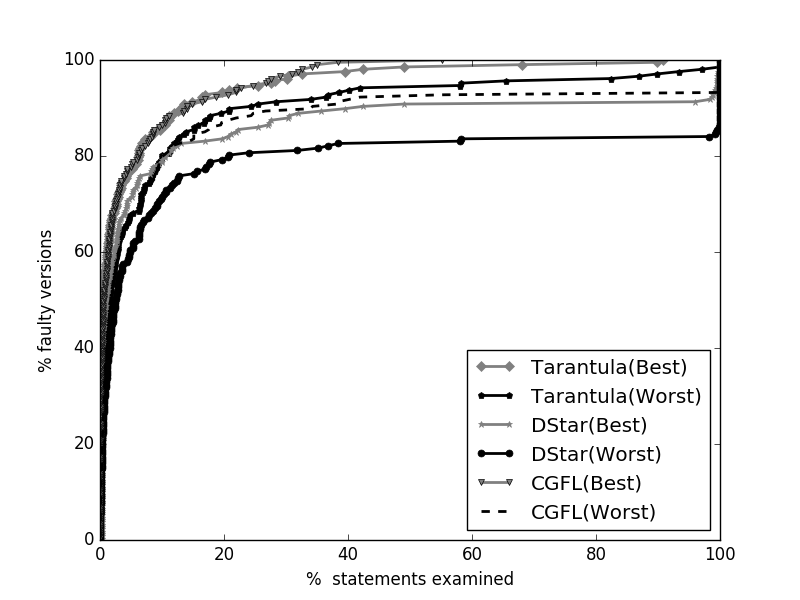}
		\caption{Effectiveness of CGFL with DStar and Tarantula for Defects4j  suite }
		\label{cgf_d4j_dt}
\end{figure}

\begin{figure}
    \centering
    \includegraphics[width=0.85\textwidth]{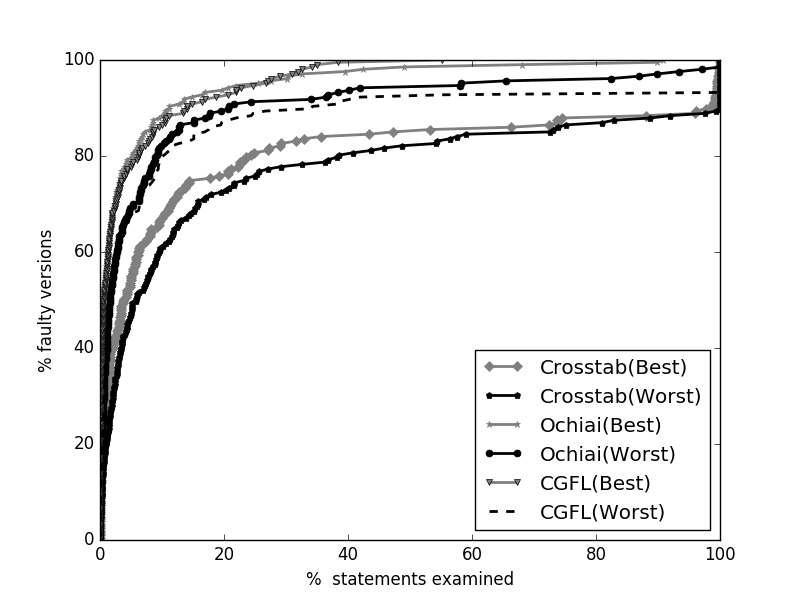}
		\caption{Effectiveness of CGFL with Crosstab and Ochiai for Defects4j  suite}
	\label{cgf_d4j_oc}
\end{figure}

\begin{figure}
    \centering
    \includegraphics[width=0.85\textwidth]{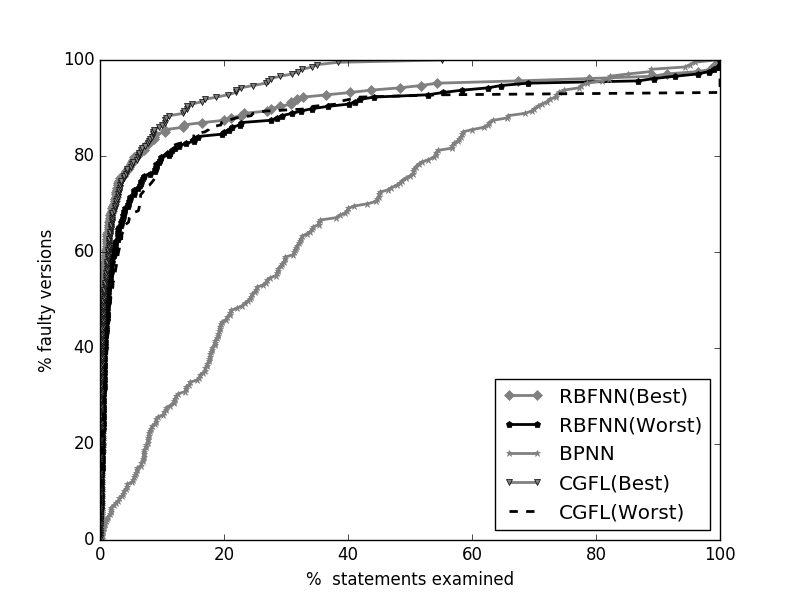}
		\caption{Effectiveness of CGFL with RBFNN and BPNN for Defects4j  suite }
		\label{cgf_d4j_rb}
\end{figure}

\begin{figure}
    \centering
    \includegraphics[width=0.85\textwidth]{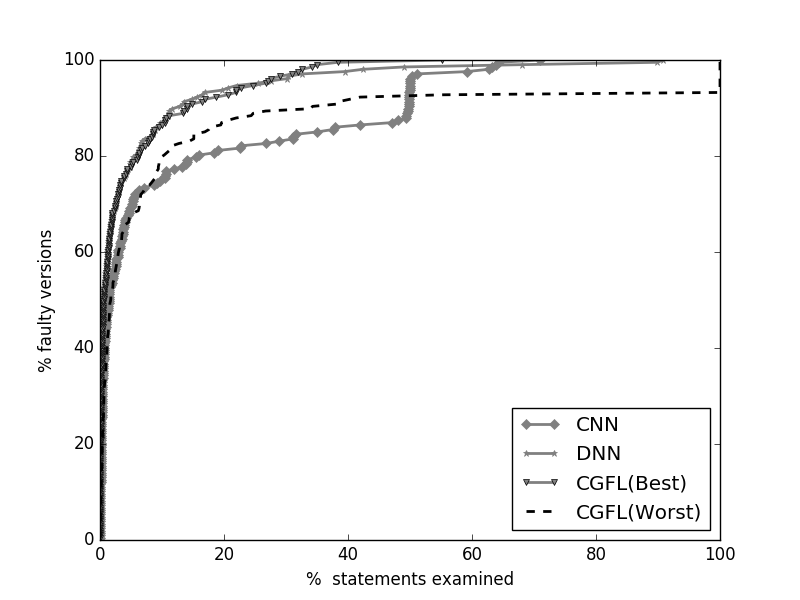}
		\caption{Effectiveness of CGFL with DNN and CNN for Defects4j  suite}
	\label{cgf_d4j_cd}
\end{figure}

\begin{figure}[]
    \centering
    \includegraphics[width=0.85\textwidth]{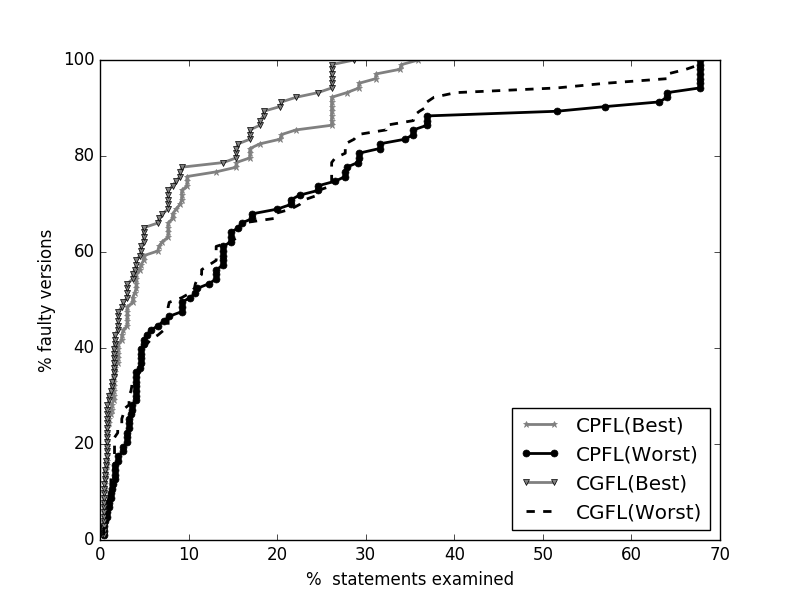}
		\caption{Effectiveness of CGFL and CPFL for Siemens  suite }
		\label{cgf_siemens_cpcg}
\end{figure}

\begin{figure}
    \centering
    \includegraphics[width=0.85\textwidth]{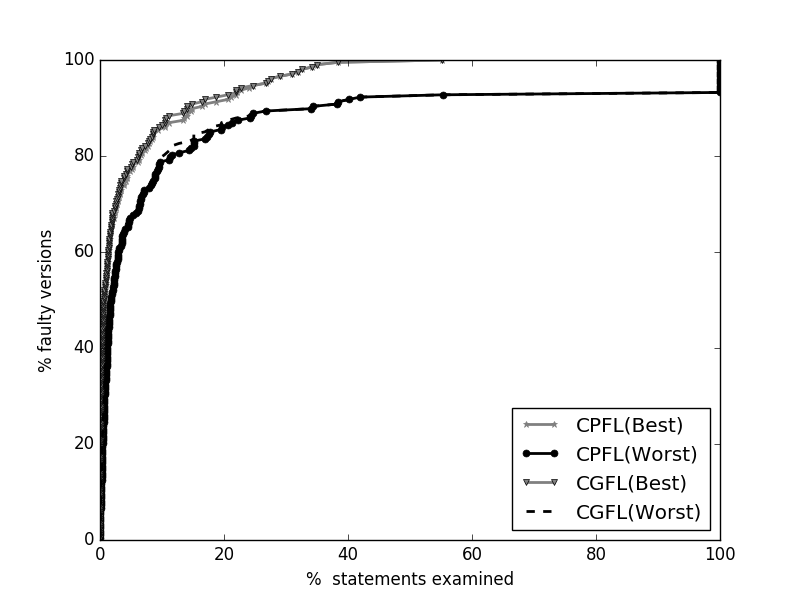}
		\caption{Effectiveness of CGFL and CPFL  for Defects4j  suite}
	\label{cgf_d4j_cpcg}
\end{figure}

 Figures \ref{cgf_siemens_dt} to \ref{cgf_d4j_cpcg} show the effectiveness comparison of CGFL and existing FL techniques. In these line graphs, the x-axis is used to denote the percentage of executable program statements examined and the y-axis  is used to depict the percentage of buggy versions localized successfully. A point (x,y) in the graph represents that the y\% of faulty programs are correctly localized by examining at most x\% of statements of the corresponding program. We represent the \textit{best} and the \textit{worst} effectiveness with two different plots in these graphs.
\par Figure \ref{cgf_siemens_dt} show the comparative results of CGFL with Tarantula and DStar using the Siemens suite data set. It can be noted from the graph that by examining only 2\% of the program statements CGFL(Best) localizes bugs in 44.66\% of faulty versions. While,  Tarantula(Best), Tarantula(Worst), DStar(Best), and DStar(Worst) localize bugs in only 19.41\%, 11.65\%, 26.21\%, and 12.62\% of faulty versions by examining the same amount of program code. On average, CGFL(Best) is 56.15\% and 50.31\% more effective than Tarantula(Best) and DStar(Best) respectively. Similarly, CGFL(Worst) is respectively 38.15\% and 34.68\% more effective than Tarantula(Worst) and DStar(Worst).
\par Figure \ref{cgf_siemens_co} pictorially represents the  comparison of the effectiveness of CFGL, Ochiai, and Crosstab using the Siemens suite. By examining only 1\% of program code, CGFL(Best) localizes bugs in 30.09\% of faulty versions whereas, Ochiai(Best) and Crosstab(Best) localize faults in only 18.44\% and 3\% of programs for the same percentage of code examination. In the worst case, CGFL(Best) is respectively 9.78\% and 45.53\% better than Ochiai(Best) and  Crosstab(Best). Similarly, CGFL(Worst) is 7.69\% and 18.46\% more effective than Ochiai(Worst) and Crosstab(Worst) in the worst case. On average, CGFL is 31.86\% and 49.16\% more effective than Ochiai and Crosstab respectively.
\par Figure \ref{cgf_siemens_rb} shows the effectiveness comparison of CGFL with BPNN and RBFNN for the Siemens suite. We can observe from the line graph in Figure \ref{cgf_siemens_rb}  that by analyzing only 0.5\% of program code, CGFL localizes bugs in 11.65\% of faulty versions. On the other hand, by examining the same amount of code RBFNN(Best) and BPNN localize bugs in only 5.8\% and 2.91\% of faulty versions respectively. In the worst case, CGFL(Best) is respectively 44.27\% and 20.55\% better than RBFNN(Best) and BPNN. Likewise,  CGFL(Worst) requires 21.54\% less code examination than RBFNN(Worst) in the worst case. Whereas, with respect to BPNN, CGFL(Worst) checks 18.45\% of more statements than BPNN in the worst-case scenario. On average, to localize bugs in all the faulty versions present in the Siemens suite, CGFL, RBFNN, and BPNN examine 11.24\%, 21.63\%, and 16.43\% of program code respectively.
\par Figure \ref{cgf_siemens_dc} shows the effectiveness comparison between CGFL, DNN, and CNN using the Siemens suite data set. We can observe from this line graph that to localize bugs in 20\% of faulty versions CGFL(Best) and CGFL(Worst) examines 0.81\%	and 1.5\% of program code. On the other hand, to locate the buggy statements in the same amount of faulty statements, DNN and CNN respectively require to examine 1.6\% and	2.04\% of executable program statements. In the worst case, CGFL(Best) is 22.85\% and	13.50\% better than DNN and CNN.  On average, CGFL is 39.28\% and 23.36\%  better than DNN and CNN respectively.	
\par Figures \ref{cgf_d4j_dt} to \ref{cgf_d4j_cd} show the comparison of results obtained using Defects4j data set for CGFL with the existing FL methods. Figure \ref{cgf_d4j_dt} shows the effectiveness comparison among Tarantula, DStar, and CGFL using a line graph. We can observe from this graph that by examining 35\% of program code, CGFL(Best) locates bugs in almost all the faulty versions. However, only 96\% of faulty versions are correctly localized by Tarantula(Best) on examining the same percentage of program code. On the other hand, while comparing with DStar, we observe that by checking only 2\% of program code, CGFL(Best) locates bugs in 68.12\% of faulty versions and DStar(Best) localizes only in 58\% of buggy programs from the same set. In the worst case, CGFL(Best) requires 35.62\% and 44.79\% of less code examination than Tarantula(Best) and DStar(Best) respectively. On average, CGFL is respectively 7.38\% and 51.1\% more effective than Tarantula and DStar for the Defects4j suite.
\par Figure \ref{cgf_d4j_oc} represents the effectiveness of CGFL, Ochiai, and Crosstab. It can be noticed from Figure \ref{cgf_d4j_oc} that CGFL has a better performance as compared to both Ochiai and Crosstab for most of the program points. By examining only 1\% of program code, CGFL(Best) and CGFL(Worst) localize bugs in 55.56\% and 37.20\% of faulty versions. On the contrary, when the same patch of program code is inspected, Crosstab(Best), Crosstab(Worst), and Ochiai(Worst) locate faults in 28.99\%, 28.29\%, and 36.71\% of faulty versions respectively. In the worst case, CGFL(Best) requires 35.62\% and 44.79\% of less code examination than Ochiai(Best) and Crosstab(Best) respectively. On average, CGFL is 61\% more effective than Crosstab. However, Ochiai requires 17\% less code examination than CGFL for the Defects4j program suite.
\par Figure \ref{cgf_d4j_rb} presents the effectiveness comparison of CGFL, BPNN, and RBFNN with plotted EXAM\_Score points in the range of 1 to 100\%. On examining 5\% of program code, CGFL(Best) and CGFL(worst) localize bugs in  77.78\% and 67.63\% of faulty versions. Whereas, BPNN and RBFNN(Worst) are able to locate bugs in only 12\% and 71\% of faulty versions. In the worst case, CGFL(Best) requires 44\% and 44.79\%  of less code examination than BPNN and RBFNN(Best). On an average, EXAM\_Score of BPNN, RBFNN(Best), RBFNN(Worst), CGFL(best) and CGFL(Worst) are 30.87\%, 8.54\%, 10.90\%, 4.40\%, and 12.03\% for Defects4j data set respectively. 
\par Figure \ref{cgf_d4j_cd} shows the comparison of CGFL, CNN, and DNN based on their effectiveness in locating program bugs. We can observe from the figure that for most of the EXAM\_Score points, CGFL(Best) performs better than both DNN and CNN. However, there are some faulty versions present for which DNN localizes the faults more effectively compared with the remaining two techniques. For the Defects4j program suite, in the worst case, CNN and DNN require 15.83\% and 35.62\% of more code examination than CGFL(Best) respectively. On average, CNN is 24\% more effective than CGFL.
\par Figures \ref{cgf_siemens_cpcg} and \ref{cgf_d4j_cpcg} show the effectiveness comparison of CGFL and CPFL over the Siemens suite and Defects4j programs respectively. CPFL is our proposed conditional probability-based fault localization approach without using the grouping method. From both figures, we can observe that for almost every program point, CGFL(Best) performs more effectively than CPFL(Best). However, there are few program points present on which CGFL(Worst) is lesser effective than CPFL(worst). But the count of such program points is comparatively low.
\par It can be observed from the Figure \ref{cgf_siemens_cpcg} that in the worst case CPFL(Best) requires 35.93\% Exam\_Score whereas CGFL(Best) examines only 28.68\% of program code. On average, CPFL(Best), CPFL(Worst), CGFL(Best), and  CGFL(Worst) require 8.40\%, 17.73\%, 6.74\%, and 15.74\% of code examination for Siemens suite respectively. Using the grouping method, we have obtained an improvement of 14.05\%, on average, over CPFL for the Siemens suite.
\par For the Defects4j program suite, we have obtained, on average, 4.97\% of improvement over CPFL(Best) by using CGFL(Best). Also, for the complete data set, we require 3\% less statements examination than the non-grouping method while localizing the faults with the CGFL method.
\par Table \ref{pwc} presents the pairwise comparison between CGFL and other fault localization techniques viz., Tarantula\cite{jones2002visualization}, DStar\cite{wong2013dstar}, Crosstab\cite{wong2008crosstab}, Ochiai\cite{naish2011model}, BPNN\cite{wong2009bp}, RBFNN\cite{wong2011effective}, CNN\cite{zhang2019cnn}, DNN\cite{zheng2016fault} and our proposed CPFL. In the sub-column titles (Best v/s Best), (Worst v/s Worst), and (Worst v/s Best), the second tag (either Best or Worst) represents the effectiveness of the existing technique. The first tag stands for the effectiveness of our proposed CGFL method. For example, the sub-column (Worst v/s Best) contains the percentage of faulty versions for which the proposed CGFL technique's worst-case behavior is more effective than the respective existing technique's best-case results. Table \ref{pwc} shows the percentage of buggy programs on which CGFL performs more effective, equally effective, and less effective than the existing fault localization techniques.
\par It is discovered from the table that CGFL(Best) is more effective in at least 58\% of the faulty versions than the existing fault localization techniques' best-case effectiveness. Also, for less than 50\% of buggy versions,  the worst-case effectiveness of CGFL  is less effective than the existing techniques' effectiveness in the worst case. Similarly, for Defects4j programs,  CGFL(Best) is at least as effective or more effective than 50\% of the faulty versions. Program on which the number of resultant tied statements (statements with the same suspiciousness score) are large with respect to an FL technique may lead to a bad performance by the CGFL method.
\begin{sidewaystable}
\caption{Pairwise comparison of CGFL with other fault localization techniques}
\label{pwc}
\scalebox{0.76}{
\centering
\begin{tabular}{|c|c|ccc|ccc|}
\hline
\multirow{2}{*}{Technique} & \multirow{2}{*}{Effectiveness} & \multicolumn{3}{c|}{Siemens}                                                               & \multicolumn{3}{c|}{Defects4j}                                                             \\ \cline{3-8} 
                           &                                & \multicolumn{1}{c|}{Best v/s Best} & \multicolumn{1}{c|}{Worst v/s Worst} & Worst v/s Best & \multicolumn{1}{c|}{Best v/s Best} & \multicolumn{1}{c|}{Worst v/s Worst} & Worst v/s Best \\ \hline
\multirow{3}{*}{Tarantula} & More effective                 & \multicolumn{1}{c|}{69.09}         & \multicolumn{1}{c|}{64.55}           & 52.73          & \multicolumn{1}{c|}{39.61}         & \multicolumn{1}{c|}{46.86}           & 28.02          \\ \cline{2-8} 
                           & Equally effective              & \multicolumn{1}{c|}{15.45}         & \multicolumn{1}{c|}{13.64}           & 4.55           & \multicolumn{1}{c|}{15.94}         & \multicolumn{1}{c|}{2.42}            & 4.35           \\ \cline{2-8} 
                           & Less effective                 & \multicolumn{1}{c|}{15.45}         & \multicolumn{1}{c|}{21.82}           & 42.73          & \multicolumn{1}{c|}{44.44}         & \multicolumn{1}{c|}{50.72}           & 67.63          \\ \hline
\multirow{3}{*}{Dstar}     & More effective                 & \multicolumn{1}{c|}{60.91}         & \multicolumn{1}{c|}{56.36}           & 40.00          & \multicolumn{1}{c|}{44.93}         & \multicolumn{1}{c|}{51.21}           & 32.85          \\ \cline{2-8} 
                           & Equally effective              & \multicolumn{1}{c|}{20.00}         & \multicolumn{1}{c|}{17.27}           & 5.45           & \multicolumn{1}{c|}{12.56}         & \multicolumn{1}{c|}{3.38}            & 4.83           \\ \cline{2-8} 
                           & Less effective                 & \multicolumn{1}{c|}{19.09}         & \multicolumn{1}{c|}{26.36}           & 54.55          & \multicolumn{1}{c|}{42.51}         & \multicolumn{1}{c|}{45.41}           & 62.32          \\ \hline
\multirow{3}{*}{Crosstab}  & More effective                 & \multicolumn{1}{c|}{76.36}         & \multicolumn{1}{c|}{75.45}           & 54.55          & \multicolumn{1}{c|}{60.87}         & \multicolumn{1}{c|}{63.29}           & 50.72          \\ \cline{2-8} 
                           & Equally effective              & \multicolumn{1}{c|}{2.73}          & \multicolumn{1}{c|}{0.91}            & 5.45           & \multicolumn{1}{c|}{8.70}          & \multicolumn{1}{c|}{2.42}            & 1.93           \\ \cline{2-8} 
                           & Less effective                 & \multicolumn{1}{c|}{20.91}         & \multicolumn{1}{c|}{23.64}           & 40.00          & \multicolumn{1}{c|}{30.43}         & \multicolumn{1}{c|}{34.30}           & 47.34          \\ \hline
\multirow{3}{*}{Ochiai}    & More effective                 & \multicolumn{1}{c|}{58.18}         & \multicolumn{1}{c|}{57.27}           & 41.82          & \multicolumn{1}{c|}{38.16}         & \multicolumn{1}{c|}{45.41}           & 27.05          \\ \cline{2-8} 
                           & Equally effective              & \multicolumn{1}{c|}{20.00}         & \multicolumn{1}{c|}{17.27}           & 3.64           & \multicolumn{1}{c|}{15.94}         & \multicolumn{1}{c|}{2.90}            & 4.35           \\ \cline{2-8} 
                           & Less effective                 & \multicolumn{1}{c|}{21.82}         & \multicolumn{1}{c|}{25.45}           & 54.55          & \multicolumn{1}{c|}{45.89}         & \multicolumn{1}{c|}{51.69}           & 68.60          \\ \hline
\multirow{3}{*}{BPNN}      & More effective                 & \multicolumn{1}{c|}{71.82}         & \multicolumn{1}{c|}{50.91}           & 50.91          & \multicolumn{1}{c|}{88.41}         & \multicolumn{1}{c|}{80.19}           & 80.19          \\ \cline{2-8} 
                           & Equally effective              & \multicolumn{1}{c|}{3.64}          & \multicolumn{1}{c|}{2.73}            & 2.73           & \multicolumn{1}{c|}{1.45}          & \multicolumn{1}{c|}{0.97}            & 0.97           \\ \cline{2-8} 
                           & Less effective                 & \multicolumn{1}{c|}{24.55}         & \multicolumn{1}{c|}{46.36}           & 46.36          & \multicolumn{1}{c|}{10.14}         & \multicolumn{1}{c|}{18.84}           & 18.84          \\ \hline
\multirow{3}{*}{RBFNN}     & More effective                 & \multicolumn{1}{c|}{64.55}         & \multicolumn{1}{c|}{60.91}           & 42.73          & \multicolumn{1}{c|}{45.41}         & \multicolumn{1}{c|}{47.83}           & 33.33          \\ \cline{2-8} 
                           & Equally effective              & \multicolumn{1}{c|}{8.18}          & \multicolumn{1}{c|}{8.18}            & 9.09           & \multicolumn{1}{c|}{12.08}         & \multicolumn{1}{c|}{1.45}            & 2.90           \\ \cline{2-8} 
                           & Less effective                 & \multicolumn{1}{c|}{27.27}         & \multicolumn{1}{c|}{30.91}           & 48.18          & \multicolumn{1}{c|}{42.51}         & \multicolumn{1}{c|}{50.72}           & 63.77          \\ \hline
\multirow{3}{*}{DNN}       & More effective                 & \multicolumn{1}{c|}{68.18}         & \multicolumn{1}{c|}{43.64}           & 43.64          & \multicolumn{1}{c|}{39.13}         & \multicolumn{1}{c|}{27.54}           & 27.54          \\ \cline{2-8} 
                           & Equally effective              & \multicolumn{1}{c|}{5.45}          & \multicolumn{1}{c|}{4.55}            & 4.55           & \multicolumn{1}{c|}{14.98}         & \multicolumn{1}{c|}{3.86}            & 3.86           \\ \cline{2-8} 
                           & Less effective                 & \multicolumn{1}{c|}{26.36}         & \multicolumn{1}{c|}{51.82}           & 51.82          & \multicolumn{1}{c|}{45.89}         & \multicolumn{1}{c|}{68.60}           & 68.60          \\ \hline
\multirow{3}{*}{CNN}       & More effective                 & \multicolumn{1}{c|}{63.64}         & \multicolumn{1}{c|}{45.45}           & 45.45          & \multicolumn{1}{c|}{55.07}         & \multicolumn{1}{c|}{39.61}           & 39.61          \\ \cline{2-8} 
                           & Equally effective              & \multicolumn{1}{c|}{11.82}         & \multicolumn{1}{c|}{3.64}            & 3.64           & \multicolumn{1}{c|}{6.28}          & \multicolumn{1}{c|}{2.90}            & 2.90           \\ \cline{2-8} 
                           & Less effective                 & \multicolumn{1}{c|}{24.55}         & \multicolumn{1}{c|}{50.91}           & 50.91          & \multicolumn{1}{c|}{38.65}         & \multicolumn{1}{c|}{57.49}           & 57.49          \\ \hline
\multirow{3}{*}{CPFL}      & More effective                 & \multicolumn{1}{c|}{30.91}         & \multicolumn{1}{c|}{30.00}           & 13.64          & \multicolumn{1}{c|}{43.48}         & \multicolumn{1}{c|}{49.76}           & 30.92          \\ \cline{2-8} 
                           & Equally effective              & \multicolumn{1}{c|}{49.09}         & \multicolumn{1}{c|}{48.18}           & 13.64          & \multicolumn{1}{c|}{14.49}         & \multicolumn{1}{c|}{1.93}            & 3.86           \\ \cline{2-8} 
                           & Less effective                 & \multicolumn{1}{c|}{20.00}         & \multicolumn{1}{c|}{21.82}           & 72.73          & \multicolumn{1}{c|}{42.03}         & \multicolumn{1}{c|}{48.31}           & 65.22          \\ \hline
\end{tabular}}
\end{sidewaystable}

\begin{sidewaystable}
\caption{Relative Improvement for CGFL wrt. existing FL techniques}
\label{RI}
\scalebox{0.8}{\begin{tabular}{|c|c|c|c|c|c|c|c|c|c|c|c|c|c|}
\hline
              & Tarantula & Tarantula & DStar & DStar & Crosstab & Crosstab & Ochiai& Ochiai & BPNN   & RBFNN & RBFNN & DNN    & CNN    \\
              &(Best)&(Worst)&(Best)&(Worst)&(Best)&(Worst)&(Best)&(Worst)&&(Best)&(Worst)&&\\\hline
Print\_Token  & 21.15           & 31.67            & 9.40        & 15.20        & 9.36           & 15.14           & 51.16        & 64.41         & 29.70  & 11.46       & 18.91        & 28.17  & 32.49  \\ \hline
Print\_Token2 & 13.27           & 19.74            & 5.49        & 8.54         & 15.82          & 22.73           & 21.88        & 31.03         & 22.12  & 51.85       & 72.58        & 25.89  & 17.16  \\ \hline
Schedule      & 17.86           & 33.82            & 4.72        & 10.13        & 34.48          & 56.10           & 52.63        & 74.19         & 22.92  & 6.90        & 14.74        & 9.24   & 14.29  \\ \hline
Schedule2     & 59.59           & 63.65            & 69.52       & 72.80        & 47.46          & 51.77           & 81.70        & 85.01         & 119.35 & 89.30       & 88.37        & 85.94  & 82.58  \\ \hline
Replace       & 40.48           & 67.38            & 57.92       & 87.94        & 32.95          & 61.57           & 57.72        & 98.38         & 40.13  & 40.57       & 75.00        & 75.02  & 90.90  \\ \hline
Tcas          & 64.55           & 72.85            & 70.97       & 76.61        & 56.21          & 68.83           & 77.55        & 79.55         & 95.00  & 44.85       & 61.41        & 91.96  & 124.00 \\ \hline
Tot\_Info     & 26.73           & 60.49            & 54.37       & 96.89        & 26.42          & 59.76           & 27.12        & 60.92         & 74.84  & 26.48       & 63.28        & 72.34  & 73.51  \\ \hline
Lang          & 67.33           & 88.15            & 19.62       & 32.76        & 16.00          & 34.99           & 78.04        & 96.06         & 14.62  & 71.26       & 127.07       & 124.94 & 53.18  \\ \hline
Math          & 101.04          & 135.25           & 22.51       & 40.26        & 15.62          & 31.12           & 111.24       & 141.75        & 15.85  & 130.78      & 156.81       & 181.89 & 65.96  \\ \hline
Mockito       & 83.14           & 142.76           & 74.66       & 91.93        & 37.11          & 111.02          & 87.06        & 146.38        & 64.68  & 36.29       & 110.73       & 179.62 & 103.14 \\ \hline
Time          & 63.82           & 106.36           & 126.55      & 110.07       & 84.28          & 129.17          & 66.33        & 109.27        & 16.93  & 71.40       & 115.38       & 107.08 & 150.28 \\ \hline
\end{tabular}}
\end{sidewaystable}
\par Table \ref{RI} presents  the relative improvement(\textit{RImp}) obtained using CGFL over the existing FL techniques as well as for the proposed CPFL method. It is calculated using Eq. 2. \textit{RImp} value shows the percentage of statements examined by CGFL with respect to any other FL technique. It can be observed from the table that for almost all the Siemens suite programs, the obtained \textit{RImp} value is lesser than 100\%. Only for programs Schedule2 and Tcas, the effectiveness of CGFL is lesser than BPNN and CNN respectively. For a few programs, CGFL requires more amount of code examination than existing techniques. The reason behind this is probably the assignment of the same suspiciousness scores to multiple statements.
On average, there is a reduction of 37.22\%,	47.32\%,	53.73\%,	22.75\%,	53.08\%,	32.48\%,	10.72\%, and 26.59\% against Tarantula, DStar, Crosstab, Ochiai, BPNN, RBFNN, DNN, and CNN respectively.

\begin{sidewaystable}
\centering
\caption{Percentage of faulty versions are successfully localized by examining Top-1\% of executable program code}
\label{top1}
\scalebox{0.73}{\renewcommand{\arraystretch}{1.5} \begin{tabular}{|c|c|c|c|c|c|c|c|c|c|c|c|c|c|c|c|}
\hline
              & \begin{tabular}[c]{@{}c@{}}Tarantula\\ (Best)\end{tabular} & \begin{tabular}[c]{@{}c@{}}Tarantula\\ (Worst)\end{tabular} & \begin{tabular}[c]{@{}c@{}}Dstar\\ (Best)\end{tabular} & \begin{tabular}[c]{@{}c@{}}Dstar\\ (Worst)\end{tabular} & \begin{tabular}[c]{@{}c@{}}Crosstab\\ (Best)\end{tabular} & \begin{tabular}[c]{@{}c@{}}Crosstab\\ (worst)\end{tabular} & \begin{tabular}[c]{@{}c@{}}Ochiai\\ (Best)\end{tabular} & \begin{tabular}[c]{@{}c@{}}Ochiai\\ (Worst)\end{tabular} & BPNN  & \begin{tabular}[c]{@{}c@{}}RBFNN\\ (Best)\end{tabular} & \begin{tabular}[c]{@{}c@{}}RBFNN\\ (Worst)\end{tabular} & DNN   & CNN   & \begin{tabular}[c]{@{}c@{}}CGFL\\ (Best)\end{tabular} & \begin{tabular}[c]{@{}c@{}}CGFL\\ (Worst)\end{tabular} \\ \hline
Print\_Token  & 40.00                                                      & 20.00                                                       & 40.00                                                  & 0.00                                                    & 0.00                                                      & 0.00                                                       & 60.00                                                   & 20.00                                                    & 0.00  & 20.00                                                  & 20.00                                                   & 20.00 & 40.00 & 60.00                                                 & 20.00                                                  \\ \hline
Print\_Token2 & 37.50                                                      & 12.50                                                       & 12.50                                                  & 0.00                                                    & 25.00                                                     & 0.00                                                       & 37.50                                                   & 12.50                                                    & 0.00  & 12.50                                                  & 0.00                                                    & 37.50 & 37.50 & 37.50                                                 & 12.50                                                  \\ \hline
Schedule      & 0.00                                                       & 0.00                                                        & 20.00                                                  & 0.00                                                    & 0.00                                                      & 0.00                                                       & 20.00                                                   & 0.00                                                     & 0.00  & 0.00                                                   & 0.00                                                    & 0.00  & 0.00  & 40.00                                                 & 0.00                                                   \\ \hline
Schedule2     & 0.00                                                       & 0.00                                                        & 0.00                                                   & 0.00                                                    & 0.00                                                      & 0.00                                                       & 0.00                                                    & 0.00                                                     & 0.00  & 0.00                                                   & 0.00                                                    & 0.00  & 0.00  & 0.00                                                  & 0.00                                                   \\ \hline
Replace       & 24.14                                                      & 13.79                                                       & 31.03                                                  & 13.79                                                   & 0.00                                                      & 0.00                                                       & 24.14                                                   & 13.79                                                    & 10.34 & 31.03                                                  & 17.24                                                   & 17.24 & 13.79 & 44.83                                                 & 20.69                                                  \\ \hline
Tcas          & 0.00                                                       & 0.00                                                        & 0.00                                                   & 0.00                                                    & 0.00                                                      & 0.00                                                       & 0.00                                                    & 0.00                                                     & 0.00  & 0.00                                                   & 0.00                                                    & 0.00  & 0.00  & 0.00                                                  & 0.00                                                   \\ \hline
Tot\_Info     & 10.53                                                      & 0.00                                                        & 21.05                                                  & 0.00                                                    & 5.26                                                      & 0.00                                                       & 15.79                                                   & 0.00                                                     & 15.79 & 15.79                                                  & 0.00                                                    & 15.79 & 5.26  & 42.11                                                 & 0.00                                                   \\ \hline
Lang          & 47.17                                                      & 26.42                                                       & 41.51                                                  & 24.53                                                   & 28.30                                                     & 20.75                                                      & 45.28                                                   & 28.30                                                    & 5.66  & 56.60                                                  & 37.74                                                   & 47.17 & 39.62 & 47.17                                                 & 28.30                                                  \\ \hline
Math          & 62.77                                                      & 37.23                                                       & 47.87                                                  & 25.53                                                   & 24.47                                                     & 11.70                                                      & 61.70                                                   & 36.17                                                    & 4.26  & 67.02                                                  & 44.68                                                   & 62.77 & 47.87 & 58.51                                                 & 32.98                                                  \\ \hline
Mockito       & 51.35                                                      & 40.54                                                       & 48.65                                                  & 40.54                                                   & 24.32                                                     & 18.92                                                      & 51.35                                                   & 43.24                                                    & 0.00  & 48.65                                                  & 45.95                                                   & 51.35 & 37.84 & 48.65                                                 & 40.54                                                  \\ \hline
Time          & 69.57                                                      & 65.22                                                       & 78.26                                                  & 73.91                                                   & 56.52                                                     & 56.52                                                      & 73.91                                                   & 69.57                                                    & 4.35  & 60.87                                                  & 56.52                                                   & 69.57 & 69.57 & 69.57                                                 & 65.22                                                  \\ \hline
\end{tabular}}
\end{sidewaystable}

\begin{sidewaystable}
\caption{Percentage of faulty versions are successfully localized by examining Top-5\% of executable program code}
\label{top5}
\centering
\scalebox{0.7}{\renewcommand{\arraystretch}{1.5} \begin{tabular}{|c|c|c|c|c|c|c|c|c|c|c|c|c|c|c|c|}
\hline
 & \begin{tabular}[c]{@{}c@{}}Tarantula\\ (Best)\end{tabular} & \begin{tabular}[c]{@{}c@{}}Tarantula\\ (Worst)\end{tabular} & \begin{tabular}[c]{@{}c@{}}Dstar\\ (Best)\end{tabular} & \begin{tabular}[c]{@{}c@{}}Dstar\\ (Worst)\end{tabular} & \begin{tabular}[c]{@{}c@{}}Crosstab\\ (Best)\end{tabular} & \begin{tabular}[c]{@{}c@{}}Crosstab\\ (worst)\end{tabular} & \begin{tabular}[c]{@{}c@{}}Ochiai\\ (Best)\end{tabular} & \begin{tabular}[c]{@{}c@{}}Ochiai\\ (Worst)\end{tabular} & BPNN & \begin{tabular}[c]{@{}c@{}}RBFNN\\ (Best)\end{tabular} & \begin{tabular}[c]{@{}c@{}}RBFNN\\ (Worst)\end{tabular} & DNN & CNN & \begin{tabular}[c]{@{}c@{}}CGFL\\ (Best)\end{tabular} & \begin{tabular}[c]{@{}c@{}}CGFL\\ (Worst)\end{tabular} \\ \hline
Print\_Token & 40.00 & 40.00 & 60.00 & 60.00 & 60.00 & 20.00 & 80.00 & 80.00 & 0.00 & 80.00 & 80.00 & 60.00 & 60.00 & 80.00 & 80.00 \\ \hline
Print\_Token2 & 62.50 & 62.50 & 37.50 & 37.50 & 62.50 & 62.50 & 62.50 & 62.50 & 25.00 & 75.00 & 75.00 & 50.00 & 50.00 & 87.50 & 87.50 \\ \hline
Schedule & 40.00 & 40.00 & 60.00 & 60.00 & 60.00 & 60.00 & 80.00 & 60.00 & 40.00 & 60.00 & 20.00 & 60.00 & 60.00 & 100.00 & 80.00 \\ \hline
Schedule2 & 0.00 & 0.00 & 12.50 & 0.00 & 0.00 & 0.00 & 12.50 & 12.50 & 12.50 & 0.00 & 0.00 & 0.00 & 12.50 & 12.50 & 0.00 \\ \hline
Replace & 48.28 & 44.83 & 65.52 & 62.07 & 51.72 & 44.83 & 65.52 & 58.62 & 27.59 & 62.07 & 62.07 & 65.52 & 68.97 & 82.76 & 65.52 \\ \hline
Tcas & 16.67 & 13.89 & 22.22 & 13.89 & 5.56 & 0.00 & 19.44 & 16.67 & 8.33 & 2.78 & 0.00 & 19.44 & 19.44 & 33.33 & 5.56 \\ \hline
Tot\_Info & 26.32 & 21.05 & 57.89 & 31.58 & 21.05 & 5.26 & 31.58 & 26.32 & 47.37 & 42.11 & 26.32 & 42.11 & 36.84 & 73.68 & 26.32 \\ \hline
Lang & 71.70 & 54.72 & 69.81 & 49.06 & 56.60 & 43.40 & 75.47 & 56.60 & 15.09 & 77.36 & 67.92 & 75.47 & 69.81 & 79.25 & 58.49 \\ \hline
Math & 80.85 & 72.34 & 69.15 & 59.57 & 47.87 & 41.49 & 82.98 & 73.40 & 7.45 & 84.04 & 75.53 & 81.91 & 68.09 & 78.72 & 71.28 \\ \hline
Mockito & 67.57 & 62.16 & 67.57 & 62.16 & 51.35 & 43.24 & 70.27 & 64.86 & 10.81 & 62.16 & 62.16 & 70.27 & 62.16 & 64.86 & 59.46 \\ \hline
Time & 86.96 & 86.96 & 86.96 & 86.96 & 82.61 & 78.26 & 86.96 & 86.96 & 21.74 & 78.26 & 78.26 & 86.96 & 82.61 & 86.96 & 86.96 \\ \hline
\end{tabular}}
\end{sidewaystable}
\par Tables \ref{top1} and \ref{top5} show the percentage of faulty versions localized by examining only Top-1\% and Top-5\% of program code respectively. The fault localization technique is considered to be better if more buggy versions are localized by just examining a smaller portion of the code. It can be observed from both Table \ref{top1} and \ref{top5} that there are different programs for which various FL techniques are unable to localize the bugs in any of the faulty  versions by examining only Top-1\% of program code. CGFL(Best) is able to locate bugs in a minimum of 37.50\% of faulty versions for all the programs except Tcas and Schedule2. On the other hand, by examining 5\% of program code, CGFL(Best) is able to locate bugs in at least 12.50\% of faulty versions for all the considered sets of programs. Moreover, for the Schedule program, CGFL(Best) is able to locate bugs in all the faulty versions by examining only Top-5\% of the code.
\begin{sidewaystable}
\caption{Time analysis of different fault localization techniques}
\label{timeAnalysis}
\scalebox{0.8}{\renewcommand{\arraystretch}{1.5} \begin{tabular}{|c|c|c|c|c|c|c|c|c|c|}
\hline
Program      & Tarantula & Dstar & Crosstab  & Ochiai & BPNN       & RBFNN       & DNN        & CNN         & CGFL \\ \hline
Print\_Token & 31ms      & 30ms  & 47ms      & 35ms   & 1min 22sec & 6min 20sec  & 3min 40sec & 5min 28sec  & 31ms \\ \hline
Pint\_Token2 & 32ms      & 34ms  & 50ms      & 33ms   & 1min 58ms  & 4min 14sec  & 4min 08sec & 6min 04sec  & 33ms \\ \hline
Replace      & 58ms      & 56ms  & 1sec 22ms & 52ms   & 3min 21sec & 11min 02sec & 8min 44sec & 12min 08sec & 58ms \\ \hline
Tcas         & 6ms       & 7ms   & 12 ms     & 8ms    & 26sec 02ms & 52sec 12ms  & 39sec 41ms & 44sec 12ms  & 7ms  \\ \hline
Tot\_info    & 6ms       & 6ms   & 10ms      & 7ms    & 1min 19ms  & 3min 18sec  & 2min 52sec & 3min 33sec  & 7ms  \\ \hline
Schedule     & 20ms      & 22ms  & 38ms      & 18ms   & 1min 28sec & 2min 05sec  & 2min 08sec & 2min 40sec  & 23ms \\ \hline
Schedule2    & 13ms      & 14ms  & 24ms      & 14ms   & 2min 12sec & 4min 27sec  & 3min 06sec & 4min 10sec  & 13ms \\ \hline
Lang     & 33sec 18ms & 31sec 02ms & 56sec 16ms & 36sec 20ms & 58sec 02ms  & 2min 12sec  & 2min 10sec  & 3min 02sec  & 30sec 12ms \\ \hline
Math     & 1sec 11ms  & 1sec 22ms  & 2sec 26ms  & 1sec 54ms  & 2min 48sec  & 4min 51sec  & 5min 10 sec & 8min 48sec  & 1sec 48ms  \\ \hline
Mockhito & 2sec 28ms  & 1sec 38ms  & 3sec 47ms  & 2sec 18ms  & 3min 12sec  & 6min 52sec  & 7min 17sec  & 9min 10sec  & 2sec 12ms  \\ \hline
Time     & 11sec 59ms & 12sec 18ms & 16sec 2ms  & 14sec 12ms & 20min 16sec & 85min 19sec & 96min 40sec & 98min 12sec & 10sec 58ms \\ \hline
\end{tabular}}
\end{sidewaystable}
\par Table \ref{timeAnalysis} presents a run time analysis of different fault localization techniques along with our proposed CGFL method.  Considered FL techniques are  Tarantula, DStar, Crosstab, Ochiai, BPNN, RBFNN, DNN, and CNN. In this table, `min' `sec', and `ms' represent minute, second, and millisecond, respectively. We have not considered the test case result generation time and statement coverage computation time because these two times were similar for all the methods. We can observe from the table that SBFL methods (Tarantula, DStar,  Crosstab, Ochiai, and CGFL) are in the order of seconds, whereas machine learning-based FL methods (CNN, DNN, BPNN, and RBFNN) require more time which is in order of minutes.   NN-based methods hold the drawback of consuming extra time due to time spent on training of the model for each faulty program. Based on the observation of results as depicted in Table \ref{timeAnalysis} it can be concluded that the CGFL method takes comparable time with SBFL techniques and shows greater efficiency than the NN-based methods.
\subsection{Threats to the validity}
\par In this section, we discuss some important threats to the validity of our proposed approach.
\begin{itemize}
    \item Construct Validity
    \begin{itemize}
      \item There can be a scenario where all the test cases may fail or all the test cases may pass, in that situation, CGFL would fail to localize the faulty statement. Therefore, the effectiveness of CGFL has a dependence on both failed and passed test cases. 
    \item While computing the suspiciousness score of statements, we have considered the same contribution of each test case. In reality, individual test cases have different contributions to deciding the suspiciousness score.
    \end{itemize}
    \item External Validity
    \begin{itemize}    
        \item Used operating systems, coverage measurement tools, compilers, and hardware platforms also have an impact on the generated results. But, to eliminate the discrepancies, we have generated results for all the existing techniques along with our proposed method on the same platform.
    \end{itemize}
    \item Internal Validity
    \begin{itemize}
        \item We assess the performance of our proposed CGFL technique on limited empirical data. There is a possibility that the technique may not work in a similar fashion with a different set of programs. But to lower the possibility of such risk, programs with varying features and from different domains are taken into consideration.
     \end{itemize}
\end{itemize}
\section{Comparison with Related Work}\label{crw}
\par In this section, we compare the proposed CGFL technique with a few related works.
\par Different slicing based techniques were reported in the literature\cite{korel1988dynamic,weiser1984program} to handle the problem of fault localization. However, these techniques are not effective enough. A number of times, slicing-based approaches return the whole program as a slice and sometimes a null set of statements too.  Furthermore, the statements do not receive any rank by using these techniques. While our technique CGFL generates a ranked list of executable statements. The ranks are assigned based on the suspiciousness of program statements to contain a bug.
\par Jones et al.\cite{jones2005empirical} developed an automated fault localization technique called Tarantula. In this method, the test execution result and statement coverage information are utilized for computing the suspicious scores of statements based on their probability of containing a bug. Their experimental result showed that Tarantula is more effective than cause-transition\cite{causetransition}, nearest neighbour\cite{setunionintersection}, set union\cite{jones2005empirical}, and set intersection\cite{jones2005empirical}  based fault localization methods. Our empirical evaluation shows that CGFL performs  \textcolor{black}{32.81\%} more effectively in locating the bugs as compared to Tarantula.
\par Renieris et al.\cite{setunionintersection} discussed the nearest neighbor technique for effective fault localization. They have defined two distance metrics viz., binary distancing and permutation distancing to calculate the similarity between a failed and a passed test case. They select an arbitrary failed test  case and compute distance with every passed test case. Subsequently, they select the passed test case with minimum distance and remove all the statements executed by that test case from the set of statements executed by the failed test case. The major drawback of this technique is the sensitivity towards the selected test  cases. If the correct pair of test cases are not selected then it would either return a null set or an irrelevant set of statements. It mainly occurs when the buggy statement is executed by both the pass and the failed test  cases. Whereas, CGFL always return a ranked list of executable statements based on their suspiciousness of containing a bug.
\par The conditional probability model adopted by Yang et al. \cite{cpfl} did not incorporate the impact of an important probability statistic $\psi_{fc}$ (probability of program element to be faulty if it has been executed by the test case) while computing the suspicious score of an executable program statement. Moreover, Yang et al.\cite{cpfl} scaled down the influence of probability parameter $\psi_{pu}$ leading to negligence of scenarios where the bug is present in conditional statements of the program. Also, the factor chosen for scaling down is the inverse of the total number of test cases while the cardinality of test cases present in the program’s test suite holds no relation to the fault's location in the program. Therefore, different numbers of test cases considered may lead to significant changes in suspicious scores calculated for the same program. CGFL is independent of the cardinality of test cases therefore, it behaves similarly for different test suites of the same program. The proposed method assigns equal weightage to all the probability statistics involved thus considering different statement invocation and execution result scenarios equally. Also, our empirical evaluation shows CGFL is 28.56\% more effective than Yang et al.\cite{cpfl} proposed approach.
\par Wong et al.\cite{wong2008crosstab} developed a statistical analysis approach for fault localization. They used the chi-square test to determine the association between the invocation of a statement and the execution outcome of a test case. Their proposed Crosstab approach is not applicable to different-sized programs. Whereas, the CGFL technique is easily applicable to any sized program. Also, we have compared the effectiveness of CGFL with Crosstab and found it to be \textcolor{black}{58.90\%} more effective than Crosstab. Along with Crosstab, we have also compared the effectiveness of CGFL with three other SBFL techniques viz., DStar\cite{wong2013dstar}, Tarantula\cite{jones2002visualization}, and Ochiai\cite{naish2011model}. Our empirical evaluation shows that CGFL requires to examine \textcolor{black}{26.84\%, 32.81\%, and 28.11\%} of less code than DStar, Tarantula, and Ochiai respectively.
\par In the past two decades, different mutation-based techniques have been used for fault localization. Some of the prominent mutation-based FL techniques are MUSE\cite{moon2014ask} and Metallaxis\cite{papadakis2015metallaxis}. Mutation-based FL techniques are effective but they are not easily applicable for large-sized programs. Since these techniques suffer from the problem of scalability.  They require huge computational power to generate a large number of mutants and thereafter to generate the statement coverage data and test execution information for each of the generated mutants. Also, it is challenging to generate all possible mutants of a program. On the other hand, our proposed CGFL technique is lightweight and straightforward. It does not require any additional investment in terms of space and time.
\par Wong et al. proposed different neural network-based FL techniques such as BPNN\cite{wong2009bp} and RBFNN\cite{wong2011effective}. Zheng et al. \cite{zheng2016fault} used a deep neural network for the same. Later, Zhang et al.\cite{zhang2017deep} extended the DNN model\cite{zheng2016fault} for FL by appending contextual information into it. Dutta et al.\cite{dutta2019hierarchically} proposed a hierarchical  approach for FL where they first localized the bug at the function level and subsequently at the statement level. Though NN-based techniques are effective they suffer from the problems of non-deterministic feedback loops and parameter estimations.  Also, the recent deep learning-based FL techniques such as DeepFL\cite{li2019deepfl} and DeepRL4FL\cite{li2021fault}  require several complex features (e.g. spectrum-based suspiciousness and complexity-based fault proneness) leading to a higher overhead for obtaining the required information.
These models also require huge training time. On the other hand, CGFL is a light-weight and simple conditional probability-based mathematical approach with minimal time requirement.

\section{Conclusion}\label{conc}
\par Software debugging is a tedious and time-consuming activity. Any improvement to it leads reduction in total software development cost. In this work, we proposed a fault localization technique that helps to mitigate the debugging cost up to a large extent. Our proposed CGFL technique is based on conditional probability-based statistics which captures the association between the execution of the statement and the test case outcome. We further appended a test case execution-based grouping approach to mitigate the ties among the statements along with more effective rank list generation. Our empirical evaluation of two popular data sets shows that on average,  CGFL requires 24.56\% less code examination than existing fault localization techniques.
\par In the future, we intend to  provide different weights to all the test cases as different test cases have different contributions in computing the suspicious score of a statement. The contribution value is computed using statement coverage information and  the execution result(pass/fail) of the test case.  In this way, the fault localizer becomes more targeted toward program faults. We also plan to incorporate statement frequency information in our CGFL technique. Statement frequency shows the number of times a statement is executed by any test case.


\bibliographystyle{elsarticle-num}
\bibliography{references}

\end{document}